\documentclass[twocolumn,aps,prc,floatfix,superscriptaddress,10pt]{revtex4-2}
\usepackage{graphicx}
\usepackage{dcolumn}
\usepackage{bm}
\usepackage[normalem]{ulem}
\usepackage{amsmath,bm}
\usepackage{amssymb}
\usepackage{longtable}
\usepackage{placeins}
\usepackage[mathlines]{lineno}
\usepackage{xcolor}
\usepackage{booktabs}
\usepackage{tabularx}
\usepackage{mathtools}
\usepackage{isotope}
\usepackage[colorlinks,linkcolor=blue,urlcolor=blue,citecolor=blue]{hyperref}
\usepackage{nccmath}
\usepackage{orcidlink}

\renewcommand{\sout}{\bgroup \color{red} \ULdepth=-.5ex \ULset}

\begin{document}

\title{Effects of light-cluster degrees of freedom on collective flows in heavy-ion collisions at FOPI energies}

\author{Xin Li \orcidlink{0009-0001-0693-6489}}
\affiliation{State Key Laboratory of Dark Matter Physics, Key Laboratory for Particle Astrophysics and Cosmology (MOE), and Shanghai Key Laboratory for Particle Physics and Cosmology, School of Physics and Astronomy, Shanghai Jiao Tong University, Shanghai 200240, China}
\affiliation{School of Physics, Henan Normal University, Xinxiang 453007, China}

\author{Si-Pei Wang \orcidlink{0009-0008-7129-8197}}
\affiliation{School of Physics, Henan Normal University, Xinxiang 453007, China}
\affiliation{State Key Laboratory of Dark Matter Physics, Key Laboratory for Particle Astrophysics and Cosmology (MOE), and Shanghai Key Laboratory for Particle Physics and Cosmology, School of Physics and Astronomy, Shanghai Jiao Tong University, Shanghai 200240, China}

\author{Rui Wang \orcidlink{0000-0002-6465-6186}}
\affiliation{%
	School of Physics, East China Normal University, Shanghai 200241, China
}%
\author{Zhen Zhang \orcidlink{0000-0003-3334-8508}}
\affiliation{Sino-French Institute of Nuclear Engineering and Technology, Sun Yat-sen University, Zhuhai 519082, China}

\author{Jie Pu \orcidlink{0000-0001-7169-4044}}
\affiliation{School of Physics, Henan Normal University, Xinxiang 453007, China}
\affiliation{Institute of Nuclear Science and Technology, Henan Academy of Science, Zhengzhou 450046, China}
\affiliation{Shanghai Research Center for Theoretical Nuclear Physics, NSFC and Fudan University, Shanghai 200438, China}

\author{Chun-Wang Ma \orcidlink{0000-0001-9372-518X}}
\affiliation{School of Physics, Henan Normal University, Xinxiang 453007, China}
\affiliation{Institute of Nuclear Science and Technology, Henan Academy of Science, Zhengzhou 450046, China}
\affiliation{Shanghai Research Center for Theoretical Nuclear Physics, NSFC and Fudan University, Shanghai 200438, China}

\author{Lie-Wen Chen \orcidlink{0000-0002-7444-0629}}
\email{Corresponding author: lwchen$@$sjtu.edu.cn}
\affiliation{State Key Laboratory of Dark Matter Physics, Key Laboratory for Particle Astrophysics and Cosmology (MOE), and Shanghai Key Laboratory for Particle Physics and Cosmology, School of Physics and Astronomy, Shanghai Jiao Tong University, Shanghai 200240, China}

\date{\today}

\begin{abstract}
	Within a lattice Boltzmann-Uehling-Uhlenbeck transport model coupled to a kinetic approach for light-cluster formation, we investigate the impact of explicit light-cluster degrees of freedom on collective flows in Au+Au collisions at FOPI energies with beam energies $E_{\rm beam}$= $120$--$1500~A$ MeV by using a density-, momentum-, and isospin-dependent N$5$LO Skyrme pseudopotential.
	We first benchmark the kinetic approach by comparing the calculated light-cluster yields with FOPI data in central Au+Au collisions.
	We then analyze the collective flows of protons and light nuclei (deuterons, tritons, $^{3}\mathrm{He}$, and $^{4}\mathrm{He}$) in mid-central collisions.
	For protons, calculations with and without dynamical light-cluster degrees of freedom are compared to quantify the influence of dynamical cluster formation on proton directed ($v_1$), elliptic ($v_2$), triangular ($v_3$), and quadrangular ($v_4$) flows.
	We find that the dynamical light-cluster effect appreciably modifies proton $v_1$--$v_4$ flows at $E_{\rm beam}=120$--$150~A$ MeV, remains visible at $E_{\rm beam}=250$--$400~A$ MeV, and gradually weakens at $E_{\rm beam}\gtrsim 600~A$ MeV. This beam-energy dependence is consistent with the decreasing abundance of light clusters at higher beam energies. For light nuclei, the kinetic approach captures the overall beam-energy dependence of the FOPI flow data, with better agreement for $E_{\rm beam}\geq 400~A$ MeV and generally overpredicted flow magnitudes at $E_{\rm beam}=120$--$250~A$ MeV. We further examine the nucleon-number scaling of $v_2/A$ in both model calculations and experimental data, finding that the kinetic light-cluster formation approach qualitatively reproduces the observed scaling behavior, with clear scaling violations around $E_{\rm beam}=120$--$400~A$ MeV and a tendency toward improved scaling at higher beam energies, although deviations remain at large scaled transverse velocity.
	These results highlight the importance of a dynamical treatment of light-cluster formation for interpreting collective flows in heavy-ion collisions below about $600~A$ MeV, although the clustering effects on proton flows are minor at higher collision energies.

\end{abstract}

\maketitle


\section{Introduction}
\label{introduction}

Heavy-ion collisions from the Fermi energy to the GeV region provide a unique laboratory for exploring the properties of strongly interacting nuclear matter under nonequilibrium conditions at densities around and above the nuclear saturation density. In this intermediate energy regime, the reaction dynamics reflects the coupled effects of mean-field evolution, nucleon-nucleon scattering, Pauli blocking, resonance excitation and decay, and cluster formation~\cite{Li:1997px,Danielewicz:2002pu,Baran:2004ih,Chen:2007fsa,Li:2008gp,Trautmann:2012nk,Ono:2018vht,Zhang:2020dvn,Tsang:2023vhh,Sorensen:2023zkk,Cozma:2025dyp,Fevre:2026rpi}.
Experimentally, collective flows, generated mainly by pressure gradients during the compressed stage and subsequently modified by rescattering and expansion, provide a sensitive observable for constraining the nuclear matter equation of state, in-medium effective interactions, and transport properties of dense nuclear matter~\cite{Stoecker:1986ci,Gale:1987zz,Li:1996ix,Pak:1997zza,Chen:1998wct,Danielewicz:1998vz,Danielewicz:1999zn,Scalone:1999mwx,Li:2000bj,Giordano:2010pv,TMEP:2016tup,Russotto:2016ucm,Nara:2020ztb,Nara:2021fuu,Li:2022wvu,TMEP:2022xjg,TMEP:2023ifw,Du:2023ype,Chen:2024aom,Wang:2024ktk,Cozma:2024cwc,Steinheimer:2024eha,Reichert:2024ayg,Wang:2024ahj,Kireyeu:2024hjo,Liu:2025pzr,Wei:2025nxe,Li:2025iqq,Zhou:2025zgn,Li:2026tjj}.

A key complication in interpreting collective-flow observables is the abundant production of light nuclei, such as deuterons ($d$), tritons ($t$), $^{3}\mathrm{He}$ ($h$), and $^{4}\mathrm{He}$ ($\alpha$), at intermediate beam energies~\cite{Ono:2018vht}. These light clusters constitute an important component of the final state, but most transport-model analyses of collective flow have focused mainly on free nucleons and pions, while light-nucleus degrees of freedom and their flow observables have not received much attention. Moreover, in many transport calculations, light clusters are not propagated as explicit degrees of freedom, since their dynamical formation, breakup, scattering, and propagation constitute a more involved problem than the transport of individual nucleons. This omission is not merely a question of particle yields: cluster formation could redistribute baryon number between free nucleons and bound state light nuclei, thereby modifying the final free-nucleon phase-space distribution and hence their collective flows, as well as the collective flows of the clusters themselves.

The treatment of light clusters in transport theory remains a nontrivial issue. In many applications, clusters are reconstructed from the final nucleon phase-space distribution by postprocessing methods, such as the coalescence approach~\cite{Sato:1981ez,Gyulassy:1982pe,Mrowczynski:1992gc,Mattiello:1996gq,Nagle:1996vp,Scheibl:1998tk,Chen:2003qj,Chen:2003ava,Chen:2004kj,Oh:2009gx,Liu:2023rlm} and the minimum spanning tree (MST) algorithm~\cite{Aichelin:1991xy,Kaur:2010ng,Goyal:2011zz,Su:2018abh,Kireyeu:2021igi,Glassel:2021rod,Liu:2022xlm}. While such approaches have achieved considerable success in describing light-nucleus observables at relativistic energies~\cite{Sun:2015jta,Zhu:2015voa,Sun:2015ulc,Sun:2016rev,Sun:2017ooe,Sun:2017xrx,Yin:2017qhg,Sun:2018jhg,Liu:2024ygk,Ma:2026wtv}, their validity at lower energies—where light-cluster abundance becomes significantly larger—remains an open question and requires further scrutiny.
In contrast, dynamical approaches treat light clusters as active degrees of freedom coupled to the transport evolution~\cite{Danielewicz:1991dh,Kuhrts:2000zs,Mohs:2020awg,Coci:2023daq,Kireyeu:2023spj,Cheng:2023rer,Wang:2023gta,Wang:2025wim}. Early work has already demonstrated that deuterons and other light fragments can be incorporated through formation and breakup channels satisfying detailed balance~\cite{Danielewicz:1991dh}. This framework has been extended to include in-medium effects, particularly Pauli blocking and the Mott dissolution of bound states, which affect cluster survival probabilities and reaction rates~\cite{Kuhrts:2000zs}. More recent kinetic approaches have generalized the description to light nuclei up to $A \le 4$ and have connected their yields in intermediate-energy heavy-ion collisions to the properties of warm dense nuclear matter~\cite{Wang:2023gta,Wang:2024raa,Wang:2025wim,Wang:2025lsm}.

Although dynamical descriptions of light-cluster formation have been substantially improved in recent years~\cite{Mohs:2020awg,Kireyeu:2023spj,Cheng:2023rer,Coci:2023daq,Wang:2023gta,Ege:2024vls,Wang:2024raa,Wang:2025lsm,Wang:2025wim,Burrello:2026abj}, the extent to which explicit light-cluster degrees of freedom modify collective-flow observables remains insufficiently quantified.
In particular, it is still unclear how strongly cluster dynamics affects the final free-proton collective flow and how this effect changes with beam energy~($E_{\rm{beam}}$). Examining the beam energy $E_{\rm{beam}}$ dependence of cluster effects is essential for identifying how the collision system evolves from a cluster-abundant regime toward a regime in which light-cluster production becomes much less prominent.
At the same time, light clusters themselves carry collective-flow information that can provide additional insight into their formation mechanism. In particular, the nucleon number scaling behavior of light-cluster flow could help test whether the collective motion of composite particles follows a simple coalescence-like picture or retains signatures of dynamical formation, breakup, and propagation in the medium. A systematic study of these effects is therefore crucial for a consistent interpretation of collective-flow observables in intermediate-energy heavy-ion collisions.

In this work, we investigate the impact of explicit light-cluster degrees of freedom on collective flows within a lattice BUU (LBUU) transport model~\cite{Wang:2018yce,Wang:2019ghr,Wang:2020ixf} supplemented by the kinetic approach for light-cluster formation~\cite{Wang:2023gta,Wang:2025wim}.
As a baseline test of the cluster-formation mechanism, we first compare the calculated light-nucleus yields with the FOPI data in central Au+Au collisions over the beam-energy range of $120$--$1500$ $A$ MeV.
We then apply the same framework to mid-central Au+Au collisions and analyze the collective flows of protons and light nuclei ($d$, $t$, $h$, and $\alpha$).
By comparing calculations with and without explicit light-cluster dynamics under otherwise identical conditions, we quantify the cluster-induced modifications of proton collective flows and examine their beam-energy dependence. We further compare the calculated proton and light-nucleus flows with available flow data over the beam-energy range from $120$ to $1500$ $A$ MeV~\cite{FOPI:2011aa} and then analyze the nucleon-number scaling behavior of elliptic flow. Our main finding is that dynamical light-cluster formation can significantly modify proton flows at lower energies and its effect on proton flows gradually weakens at $E_{\rm beam}\geq 600~A$ MeV.

This paper is organized as follows. In Sec.~\ref{Methodology}, we introduce the LBUU transport framework, the dynamical treatment of light clusters, and the basic concept of anisotropic flow. In Sec.~\ref{discussion}, we present the yield benchmark, analyze the effects of explicit light-cluster degrees of freedom on proton flows, as well as the light-nucleus flows described by the kinetic approach, and discuss the nucleon-number scaling of elliptic flow. A summary and outlook are given in Sec.~\ref{summary}.

\section{Theoretical framework}
\label{Methodology}

\subsection{The lattice BUU transport model and dynamical formation of light clusters}
In this work, we employ the LBUU transport model~\cite{Wang:2018yce,Wang:2019ghr,Wang:2020ixf} to study proton and light-cluster formation in the Au+Au collisions at FOPI energies over a beam-energy range of $120$--$1500$ $A$ MeV. In the kinetic approach employed in the present study, light clusters, as for nucleons, pions, $\Delta$ resonances and nucleon resonances, are included dynamically in the kinetic equations that are derived from the real-time Green's function formalism~\cite{Danielewicz:1991dh}. These kinetic equations, which govern the
time evolution of their Wigner functions or phase-space distribution function $f_{\tau}(\vec{r}, \vec{p}, t)$, can be written as~\cite{Wang:2023gta,Wang:2025wim}
\begin{equation}\label{E:KE}
	(\partial_t +{\boldsymbol\nabla}_p\epsilon_i\cdot{\boldsymbol\nabla}_r -{\boldsymbol\nabla}_r\epsilon_i\cdot{\boldsymbol\nabla}_p)f_i = I_i^{\rm coll}[f_n,f_p,f_d,\cdots],
\end{equation}
with $i$ denoting the proton ($p$), neutron ($n$), deuteron ($d$), triton ($t$), $^{3}\mathrm{He}$ ($h$) and $^{4}\mathrm{He}$ ($\alpha$), as well as the different charged states of pion ($\pi$), $\Delta$ resonances and nucleon resonances. In the above equation, $\epsilon_i[f_n,f_p,\cdots]$ represents the single-particle energy, which can be derived from an energy-density functional, and $I_i^{\rm coll}$ is the collision integral.

In the present study, the energy-density functional is constructed based on the N5LO Skyrme pseudopotential~\cite{Wang:2023zcj,Wang:2024xzq,Li:2025myg}.
We adopt the maximum-a-posteriori (MAP) parameter set obtained from a Bayesian
inference on proton flows at HADES energies~\cite{Li:inprep}. Following Refs.~\cite{Wang:2023zcj,Wang:2024xzq,Li:2025myg}, the properties of symmetric nuclear matter at the saturation density $\rho_0$ are characterized by the binding energy per nucleon $E_0(\rho_0)$, the incompressibility coefficient $K_0$, the skewness coefficient $J_0$, and the kurtosis coefficient $I_0$; the density dependence of the nuclear symmetry energy is characterized by the symmetry energy at saturation $E_{\rm sym}(\rho_0)$, the slope coefficient $L$, the curvature coefficient $K_{\rm sym}$, the skewness coefficient $J_{\rm sym}$, and the kurtosis coefficient $I_{\rm sym}$; and the momentum dependence of the symmetry potential is additionally characterized by the isospin splitting coefficient of the nucleon effective mass $\Delta m_1^*(\rho_0)$. The set of macroscopic quantities further includes the gradient-term parameter $E^{[2]}$ (defined as in Refs.~\cite{Wang:2019ghr,Wang:2024xzq,Li:2025myg}), which is set to $E^{[2]} = -310~\mathrm{MeV}\,\mathrm{fm}^{5}$ to reproduce the experimental ground-state binding energy of $^{197}\mathrm{Au}$ within the Thomas--Fermi method~\cite{Wang:2019ghr,Wang:2020ixf}.
In addition, for the isoscalar momentum-dependent mean-field potential, it is fixed from fitting the nucleon optical model potential (and its extrapolation above $1$~GeV) in symmetric nuclear matter at nuclear saturation density obtained by Hama et al.~\cite{Hama:1990vr,Cooper:1993nx} (see, e.g., Refs.~\cite{Xu:2014cwa,Wang:2025uyl,Li:2025uku} for the details ). 
In such a way, once the macroscopic quantities [i.e., $\rho_0$, $E_0(\rho_0)$, $K_0$, $J_0$, $I_0$, $E_{{\mathrm{sym}}}(\rho_0)$, $L$, $K_{{\mathrm{sym}}}$, $J_{{\mathrm{sym}}}$, $I_{{\mathrm{sym}}}$, $\Delta m_{1}^*(\rho_0)$, and $E^{[2]}$] are specified, all interaction parameters of the N5LO pseudopotential can then be determined~\cite{Wang:2019ghr,Wang:2024xzq,Li:2025myg}.
The adopted values of these macroscopic quantities are summarized in Table~\ref{tab:eos-parameters}.
\begin{table}[t]
	\centering
	\fontsize{9}{11}\selectfont
	\setlength{\tabcolsep}{7pt} 
	\caption{The macroscopic parameters of N$5$LO Skyrme pseudopotential used in this work.}
	\label{tab:eos-parameters}
	\begin{tabular}{lclc}
		\toprule
		\toprule
		Parameters & Value & Parameters & Value \\
		\midrule
		$\rho_{0}$ ($\mathrm{fm}^{-3}$)
		& $0.160$
		& $E_{\rm sym}(\rho_0)$ ($\mathrm{MeV}$)
		& $30.3$
		\\
		$E_0(\rho_0)$ ($\mathrm{MeV}$)
		& $-16.0$
		& $L$ ($\mathrm{MeV}$)
		& $30.1$
		\\
		$K_0$ ($\mathrm{MeV}$)
		& $235.0$
		& $K_{\rm sym}$ ($\mathrm{MeV}$)
		& $-251.5$
		\\
		$J_0$ ($\mathrm{MeV}$)
		& $-195.0$
		& $J_{\rm sym}$ ($\mathrm{MeV}$)
		& $397.1$
		\\
		$I_0$ ($\mathrm{MeV}$)
		& $1611.8$
		& $I_{\rm sym}$ ($\mathrm{MeV}$)
		& $-1994.7$
		\\
		$E^{[2]}$ (${\rm MeV\cdot fm}^5$)
		& $-310$
		&$\Delta m_{1}^*(\rho_0)$
		& $0.39$
		\\
		\bottomrule
		\bottomrule
	\end{tabular}
\end{table}

The corresponding single-particle potentials are specified consistently with this extended Skyrme energy-density functional based on the N5LO pseudopotential.
For nucleons, $\epsilon_i$ contains the density-, momentum- and isospin-dependent mean-field potential derived from the functional.
The potentials of $\Delta$ resonances are taken as linear combinations of the neutron and proton potentials~\cite{UmaMaheswari:1997mc,Li:2025myg}, while the single-particle potentials of pions and higher baryon resonances ($\Delta^{*}$ and $N^{*}$) are neglected~\cite{Li:2025uku,Li:2025myg}.
For light clusters, the single-particle potential is constructed as the sum of the single-particle potentials of their constituent nucleons~\cite{Wang:2025lsm,Wang:2025wim}.

The collision integral $I_i^{\rm coll}$ on the right-hand side of Eq.~(\ref{E:KE}) consists of a gain term ($<$) and a loss term ($>$),
\begin{equation}\label{E:Ic}
	I_i^{\rm coll} = K_i^{<}[f_n,f_p,\cdots](1\pm f_i) - K_i^{>}[f_n,f_p,\cdots]f_i,
\end{equation}
where the plus and minus signs are for bosons and fermions, respectively. Both gain and loss terms contain contributions from various scattering channels, which can be obtained from diagrammatic expansions of many-body Green's functions~\cite{Danielewicz:1991dh}.
In the present work, the collision integral includes both elastic and inelastic scattering processes. The free nucleon-nucleon elastic cross sections are taken from the parametrization of experimental scattering data~\cite{Cugnon:1996kh}, while their in-medium modifications are implemented following the thermodynamic $T$-matrix-based prescription used in Refs.~\cite{Wang:2020xgk,Li:2025myg}.
The overall strength of the in-medium effect is controlled by a free parameter $\alpha_{NN}$ (see Refs.~\cite{Wang:2020xgk,Li:2025myg} for the detailed expression), whose value is set to the MAP value $\alpha_{NN}=2.13$ obtained from the same Bayesian inference~\cite{Li:inprep}.

For elastic scatterings involving baryons other than two nucleons, we follow the standard treatment adopted in transport-model comparisons~\cite{Kolomeitsev:2004np,TMEP:2016tup,TMEP:2017mex,TMEP:2021ljz,TMEP:2022xjg,TMEP:2019yci,TMEP:2023ifw} and use an isotropic constant cross section~\cite{Li:2025myg}. Elastic scatterings involving light clusters are also included consistently in the collision integral. For nucleon--cluster and cluster--cluster elastic channels, $N\nu \rightarrow N\nu$ and $\nu\nu' \rightarrow \nu\nu'$ with $\nu,\nu'=d,t,h,\alpha$, the corresponding cross sections are constructed by scaling the parametrized $Nd\rightarrow Nd$ elastic cross section~\cite{Oh:2009gx,Wang:2023gta,Wang:2025wim}.

For the inelastic scattering part including two-body processes, a parameterized isotropic cross section for the $NN \rightarrow N\Delta$ scattering as in Ref.~\cite{Li:2025uku} is employed. For the $N\Delta \rightarrow NN$ scattering, its cross section is related to that of $NN \rightarrow N\Delta$ by the detailed balance condition (see Ref.~\cite{TMEP:2023ifw} for more details). The resonance decay and productions, i.e., $\Delta(N^*,\Delta^*) \leftrightarrow N\pi$ and $\Delta^*(N^*) \leftrightarrow \Delta\pi$ are included with the cross sections and decay width taken from Refs.~\cite{TMEP:2023ifw,SMASH:2016zqf}.

For inelastic reactions involving light nuclei, we include the following nucleon-induced catalytic reactions $NNN\leftrightarrow Nd$, $NNd\leftrightarrow Nt(h)$, $NNNN\leftrightarrow Nt(h)$, $NNNNN\leftrightarrow N\alpha$, $NNNd\leftrightarrow N\alpha$, $NNt(h)\leftrightarrow N\alpha$, and the two-body inelastic channel $N\alpha\leftrightarrow dt(h)$.
In addition, pion-catalyzed production and dissociation of light nuclei are mediated by an intermediate $\Delta$ resonance. The corresponding reaction channels are $\Delta N\leftrightarrow \pi d$, $\Delta NN\leftrightarrow \pi t$, $\Delta NN\leftrightarrow \pi h$, and $\Delta NNN\leftrightarrow \pi \alpha$.
The cross sections for these catalytic reactions are evaluated within the impulse approximation~\cite{Wang:2023gta,Sun:2022xjr}, in which the corresponding transition matrix elements are factorized into the light-nucleus internal momentum-space wave functions and the elementary nucleon-nucleon elastic-scattering amplitudes, with normalizations constrained by measured nucleon-nucleus breakup cross sections~\cite{Wang:2023gta}.

The above kinetic equations can be solved using the test-particle method~\cite{Wong:1982zzb}, with the drift terms on the left-hand side of Eq.~(\ref{E:KE}) treated by the lattice Hamiltonian method~\cite{Lenk:1989zz,Wang:2019ghr}, and the collision integral on the right-hand side of Eq.~(\ref{E:KE}) treated in a stochastic approach~\cite{Danielewicz:1991dh,Wang:2020xgk,Wang:2020ixf}.
The initial ground states of the colliding nuclei are prepared with the Thomas-Fermi initialization~\cite{Wang:2019ghr,Wang:2020ixf}, which ensures the stability of the ground state evolution~\cite{Wang:2019ghr,Li:2025myg}.

As for light clusters, their formation in the above reaction channels is further constrained by in-medium effects. In the nuclear medium, the binding of a light cluster could be weakened by Pauli blocking from surrounding nucleons. From the in-medium Schr\"odinger equation, the binding energy $E_{\rm B}(\mathbf{P})$ of a light cluster with center-of-mass momentum $\mathbf{P}$ may vanish when the nucleon phase-space occupation around the cluster becomes sufficiently large, leading to the dissolution of the bound state, called Mott effect~\cite{Ropke:1982ino,Ropke:1983lbc}. This suppression is therefore essential for a proper dynamical description of light-cluster production in heavy-ion collisions. To incorporate this effect in transport simulations, we employ the phase-space excluded-volume approach introduced in Ref.~\cite{Wang:2025lsm}. This approach captures the essential Pauli-blocking mechanism of the in-medium Schr\"odinger equation while requiring significantly less computational effort. Specifically, the formation of a light cluster of species $\nu$ with mass number $A$ and momentum $\mathbf{P}$ is allowed only when the averaged phase-space occupation $\langle f_\tau \rangle_\nu(\mathbf{P})$ of the surrounding nucleons is below a cutoff value $F_A^{\rm cut}$,
\begin{equation}
	\langle f_\tau \rangle_\nu(\mathbf{P}) \equiv \int f_\tau^{\rm tot}(\mathbf{p})|\tilde{\phi}_{\nu,\mathbf{P}}({\bf p})|^2\frac{{\rm d}{\bf p}}{(2\pi\hbar)^3} \leqslant F^{\rm cut}_A .
	\label{E:fcut}
\end{equation}
Here, $\tau=n$ or $p$, and $|\tilde{\phi}_{\nu,\mathbf{P}}({\bf p})|^2$ denotes the normalized one-body momentum distribution of the nucleons inside the light cluster $\nu$. The parameters $F_A^{\rm cut}$ can be regarded as surrogates of the strength of the Mott effect, with smaller values of $F_A^{\rm cut}$ corresponding to a stronger Mott effect, i.e., making it
easier for light clusters to dissolve in nuclear medium.
The total nucleon occupation $f_\tau^{\rm tot}$ includes contributions from both unbound nucleons and nucleons bound in light clusters~\cite{Wang:2025lsm}, differing from the original phase-space excluded-volume implementation~\cite{Danielewicz:1991dh,Kuhrts:2000zs,Wang:2023gta}, where only unbound nucleons are included in the medium occupation.

In heavy-ion collision simulations, the above criterion for light-cluster formation is incorporated into the kinetic approach through the collision integral in Eq.~(\ref{E:KE}), i.e., light clusters produced through many-body scattering channels are allowed only when Eq.~(\ref{E:fcut}) is satisfied in the corresponding spatial cell and time step. Through Eq.~(\ref{E:KE}), the Mott effect manifests itself in heavy-ion collisions, with its strength characterized by the cutoff parameters $\mathbf{F}^{\rm cut}\equiv(F_2^{\rm cut},F_3^{\rm cut},F_4^{\rm cut})$. In this work, we use the posterior most probable values $\mathbf{F}^{\rm cut}=(0.192,0.248,0.345)$ obtained in Ref.~\cite{Wang:2025wim} from Bayesian inference based on measured light-nucleus yields to estimate the strength of the Mott effect.

Furthermore, we perform two sets of calculations to quantify the influence of explicit light-cluster degrees of freedom on collective flows. One includes light clusters ($d$, $t$, $h$, and $\alpha$) as dynamical degrees of freedom with the treatments described above, while the other switches off the light-cluster degrees of freedom and removes all cluster-related channels from the collision integral.
Additionally, in the numerical implementation, the lattice spacing is set to $1$ fm and the number of test particles is taken to be $30{,}000$.
The reaction is evolved with a time step of $0.2~{\rm fm}/c$ up to a beam-energy-dependent stopped time $t_{\rm{stop}}$ chosen to ensure convergence of the relevant observables: $t_{\rm{stop}}=300~{\rm fm}/c$ for $120$ and $150~A$ MeV, $t_{\rm{stop}}=200~{\rm fm}/c$ for $250$--$600~A$ MeV, $t_{\rm{stop}}=150~{\rm fm}/c$ for $800~A$ MeV, $t_{\rm{stop}}=100~{\rm fm}/c$ for $1000~A$ MeV, and $t_{\rm{stop}}=80~{\rm fm}/c$ for $1200$ and $1500~A$ MeV.

\subsection{Anisotropic flows and centrality selection}

The anisotropic flow coefficients $v_n$ quantify the momentum anisotropy and are defined as the Fourier components in the expansion of the particle transverse momentum $p_t$ spectra with respect to the azimuthal angle $\phi$ relative to the reaction plane~\cite{Poskanzer:1998yz}:
\begin{equation}
	\label{eq:flow_n}
	\small
	E \frac{\mathrm{d}^3 N}{\mathrm{d} p^3}= \frac{1}{2\pi}\frac{\mathrm{d}^{2} N}{p_{t} \mathrm{d} p_{t} \mathrm{d} y } \left[ 1 + \sum_{n=1}^{\infty} 2 v_{n} (p_{t},y) \cos{(n \phi)} \right].
\end{equation}
The anisotropic flow coefficients $v_n$ can be expressed as particle number averages in terms of single-particle momentum components. In the present work, we focus on the directed flow $v_1$, elliptic flow $v_2$, triangular flow $v_3$, and quadrangular flow $v_4$ which are evaluated from the final-state particle momenta as~\cite{Chen:2004dv,Chen:2004vha}
\begin{align}
	\label{eq:flow_expressions}
	v_{1}(p_{t}) &= \left \langle \frac{p_{x}}{p_{t}} \right \rangle, \\
	v_{2}(p_{t}) &= \left \langle \frac{p_{x}^{2}-p_{y}^{2}}{p_{t}^{2}} \right \rangle, \\
	v_{3}(p_{t}) &= \left \langle \frac{p_{x}^{3}- 3 p_{x} p_{y}^{2}}{p_{t}^{3}} \right \rangle, \\
	v_{4}(p_{t}) &= \left \langle \frac{p_{x}^{4}- 6 p_{x}^{2} p_{y}^{2} + p_{y}^{4} }{p_{t}^{4}} \right \rangle,
\end{align}
with $p_x$ and $p_y$ denoting the momentum components parallel and perpendicular to the reaction plane, respectively. The brackets represent an average over particles in the specified rapidity and transverse-momentum interval.

Since collective flows depend strongly on the collision geometry, a consistent treatment of centrality is required when comparing transport-model calculations with experimental data. In the FOPI analysis, the collision centrality selection was obtained by binning distributions of either the detected charged-particle multiplicity (MUL) or the ratio of the total transverse to longitudinal kinetic energies in the center-of-mass frame (ERAT). The corresponding impact-parameter intervals were inferred from the measured differential cross sections of the ERAT or multiplicity distributions using a geometrical sharp-cut approximation~\cite{FOPI:2011aa}.

In this work, the centrality is characterized by the interval of the scaled impact parameter $b_0$, which is defined as $b_0=b/b_{\rm max}$, where $b$ is the impact parameter and $b_{\rm max}=13.38$~fm for Au+Au collisions. We use the impact-parameter interval corresponding to the specified FOPI centrality class and calculate the final observable as a weighted average over simulations performed at several representative impact parameters (see Ref.~\cite{Li:2025myg} and references therein for details). For the centrality range $0.25<b_0<0.45$, the corresponding impact parameter ranges from $3.3$ to $6.0$~fm. This procedure enables a consistent comparison between the theoretical results and the FOPI data within the same specified impact-parameter interval.

\section{Results and discussion}

\label{discussion}
In this section, we present the simulation results from the LBUU transport model incorporating the dynamical light-cluster formation mechanism described above.
We first benchmark the kinetic approach for light-nucleus formation by comparing the calculated light-nucleus yields with FOPI data~\cite{FOPI:2010xrt} in central ($b_{0}<0.15$) Au+Au collisions. We then examine the beam-energy dependence of proton directed ($v_1$), elliptic ($v_2$), triangular ($v_3$), and quadrangular ($v_4$) in mid-central ($0.25<b_{0}<0.45$) Au+Au collisions over the beam-energy range from $120$ to $1500~A$ MeV, in order to quantify how the explicit inclusion of light-cluster degrees of freedom affects proton collective flows. We also compare the calculated flows of protons and light nuclei with the available FOPI data~\cite{FOPI:2011aa} over the same energy range. Finally, we discuss the nucleon-number scaling behavior of light-nucleus flows and its implications for the dynamical origin of cluster formation in intermediate-energy heavy-ion collisions.

\subsection{Benchmark of the kinetic approach with light-nucleus yields}
\label{LNyield}
\begin{figure}[htbp]
	\centering
	\includegraphics[width=0.45\textwidth]{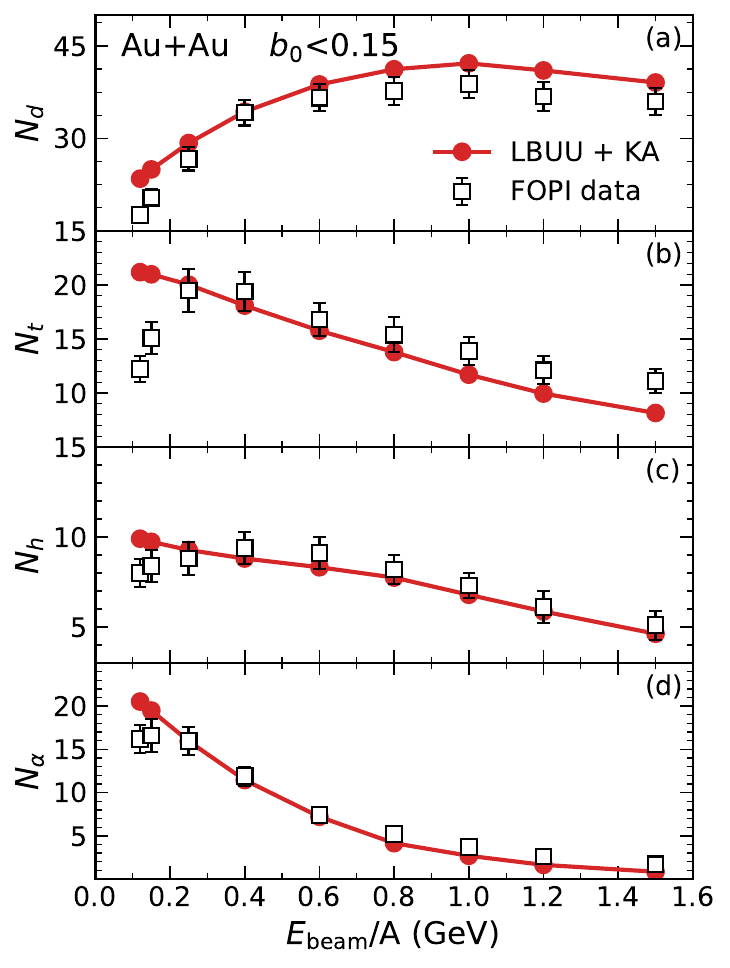}
	\caption{Beam-energy dependence of the yields of deuterons ($d$), tritons ($t$), $^3$He ($h$), and $^4$He ($\alpha$) particles in  Au+Au collisions with $b_0<0.15$. The LBUU calculations with kinetic approach (KA) for light-cluster formation are compared with the FOPI data~\cite{FOPI:2010xrt}.}
	\label{fig:FOPILNyield}
\end{figure}

Before discussing collective-flow observables, we first examine whether the dynamical light-cluster treatment provides a reasonable description of the light-nucleus yields in the FOPI energy range of $120$--$1500~A$ MeV. Figure~\ref{fig:FOPILNyield} shows the beam-energy dependence of the multiplicities of deuterons, tritons, $^{3}\mathrm{He}$, and $^{4}\mathrm{He}$ in central Au+Au collisions (impact parameter $b=1.4$~fm, corresponding to a centrality selection $b_0<0.15$). The LBUU calculations with the kinetic approach (KA) for light-nucleus formation are compared with the corresponding FOPI data~\cite{FOPI:2010xrt}.

One can see from Fig.~\ref{fig:FOPILNyield} that the calculated results give a reasonable description of the measured light-nucleus yields over the beam-energy range from $120$ to $1500$ $A$ MeV. However, some visible deviations remain at lower beam energies. In particular, at $E_{\rm beam}=120$ and $150$ $A$ MeV, the calculated multiplicities of light nuclei are systematically higher than the FOPI data. This overestimation is most evident for tritons and $\alpha$ particles, and is also present, though less pronounced, for deuterons and $^3$He (a discussion on the lower-energy deviation will be given below). With increasing beam energy, for example around $250$--$600$ $A$ MeV, the discrepancy becomes much smaller and the calculation follows the measured yields more closely. At higher beam energies, from $800$ to $1500~A$ MeV, the calculation reproduces the decreasing trend of the measured light-nucleus yields and describes the FOPI data reasonably, consistent with the reduced survival probability of light clusters at higher collision energies. In addition, pion-catalyzed reactions are expected to become increasingly important in this energy regime. These reactions are included in our present calculations as mentioned before, which may contribute to the improved agreement with the data compared with Ref.~\cite{Wang:2023gta}, where such reactions were not taken into account.

Despite the remaining low-energy deviations, the overall agreement with the FOPI data on light-nucleus yields indicates that the kinetic approach for light-nucleus formation captures the main features of cluster formation, breakup, rescattering, and in-medium suppression in this energy region. This provides a useful basis for the following flow analysis. Since the abundance of light clusters reflects how much baryon number is redistributed from the free-nucleon sector into bound clusters, the beam-energy dependence of the yields provides a direct baseline for understanding the cluster-induced modification on other nucleon observables. The decreasing abundance of light clusters at higher beam energies therefore naturally suggests a weaker influence of light-cluster degrees of freedom on proton flow observables, which will be examined in the next subsection.

In addition, we would like to point out that the comparison at low beam energies (i.e., $E_{\rm beam}=120$ and $150$ $A$ MeV) requires a careful interpretation. The calculations tend to overestimate the yields of some light clusters. This deviation may indicate that, in this low-energy region, light-nucleus production is not governed solely by the kinetic formation and breakup channels of clusters with $A\leq 4$.
At such low beam energies, contributions from heavier fragments and residual nuclei may become more important. Correlations among light clusters may also promote their aggregation into heavier fragments and thereby affect the calculated light-cluster yields~\cite{Ono_2013,Ono:2016xun,Ono:2018vht}. Since neither heavier fragments nor such inter-cluster correlations are included in the present calculation, part of the baryon number that would otherwise be bound in heavier fragments may remain available for producing light clusters.
Moreover, the predicted yields at these lower energies are particularly sensitive to the strength of the in-medium suppression implemented through the cutoff parameters $F_A^{\rm cut}$. A weaker effective suppression at high phase-space occupation could enhance the survival probability of light clusters and lead to an overestimate of their final multiplicities. Because light clusters are most abundant at the low-energy end of the studied range, such uncertainties in the formation, dissociation, and in-medium survival probabilities are amplified in the calculated multiplicities.

\subsection{Beam-energy dependence of proton collective flows}
\label{energyflow}

\begin{figure}[htbp]
	\centering
	\includegraphics[width=0.45\textwidth]{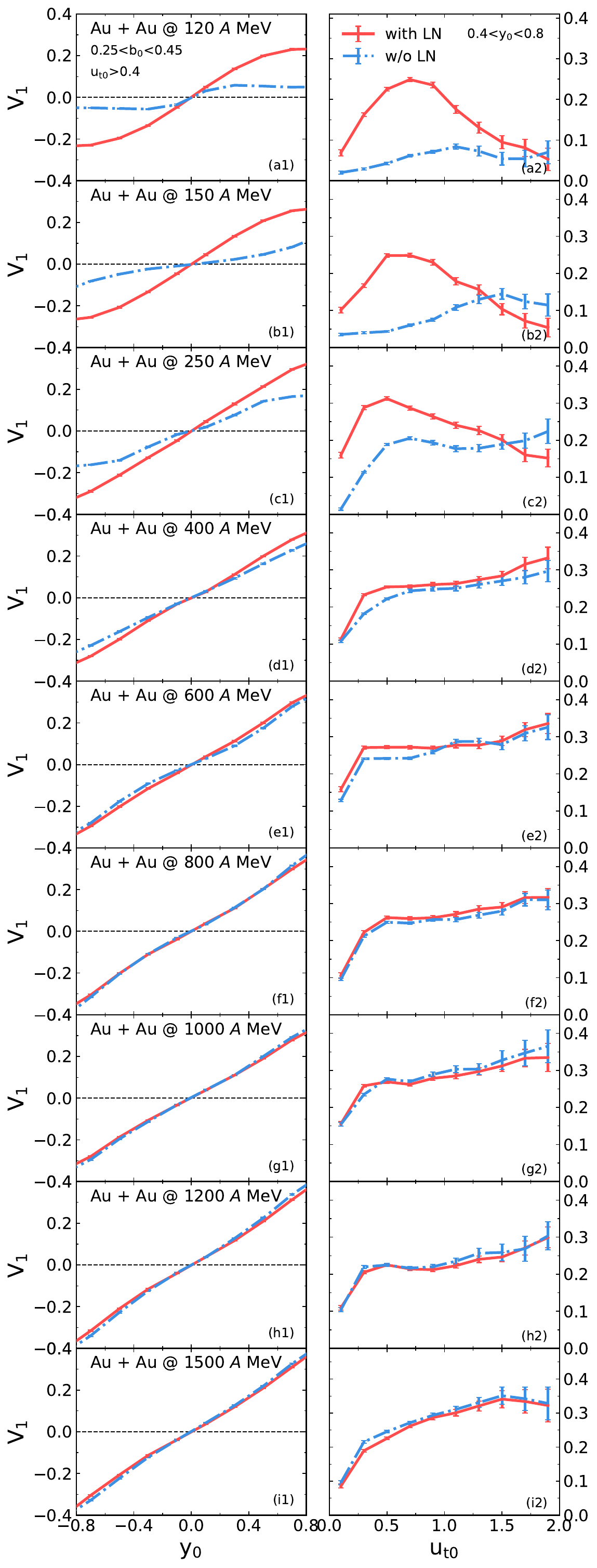}
	\caption{Directed flow $v_1$ of protons in Au+Au collisions at $E_{\rm beam}=120-1500~A$ MeV with $0.25<b_{0}<0.45$. Left: $v_1(y_0)$ for $u_{t0}>0.4$. Right: $v_1(u_{t0})$ for $0.4<y_0<0.8$. Red solid and blue dash-dotted lines show calculations with and without light nuclei, respectively.}
	\label{fig:v1energy}
\end{figure}

\begin{figure}[htbp]
	\centering
	\includegraphics[width=0.48\textwidth]{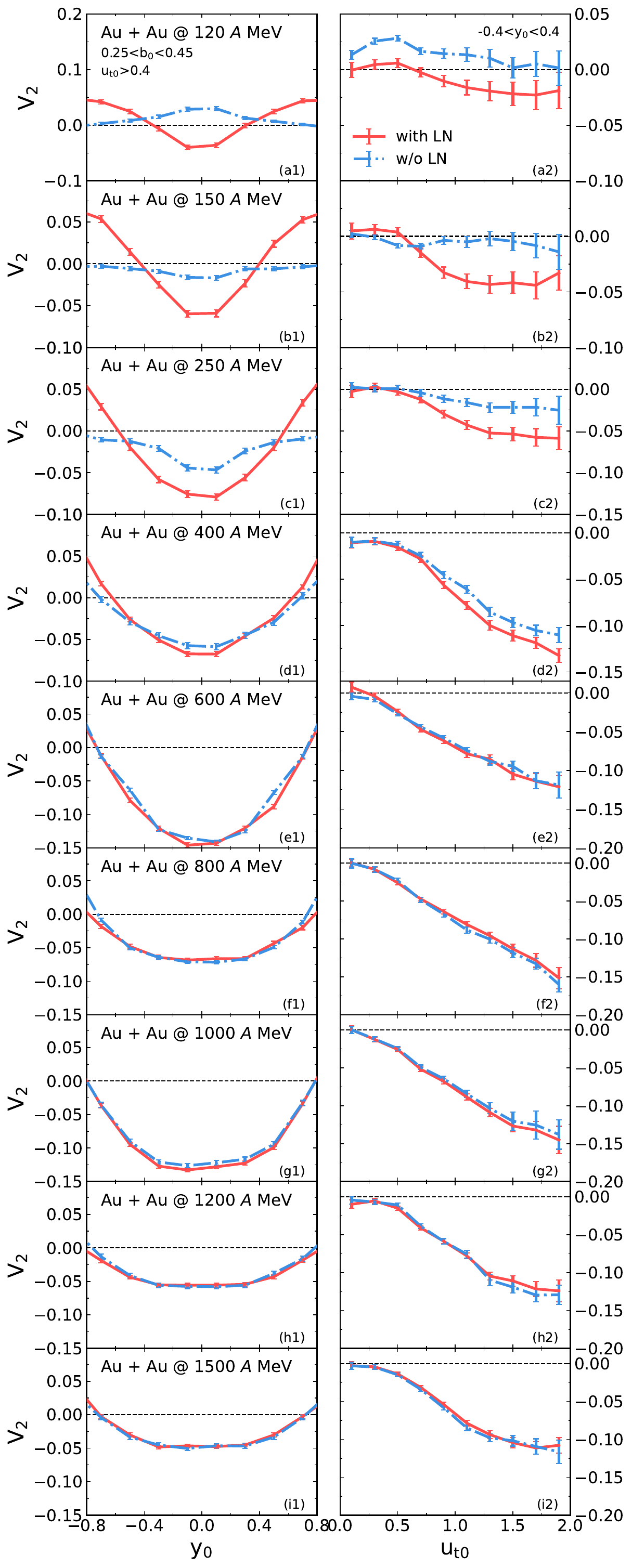}
	\caption{Elliptic flow $v_2$ of protons in Au+Au collisions at $E_{\rm beam}=120-1500~A$ MeV with $0.25<b_{0}<0.45$. Left: $v_2(y_0)$ for $u_{t0}>0.4$. Right: $v_2(u_{t0})$ for $-0.4<y_0<0.4$. Red solid and blue dash-dotted lines show calculations with and without light nuclei, respectively.}
	\label{fig:v2energy}
\end{figure}

\begin{figure}[htbp]
	\centering
	\includegraphics[width=0.48\textwidth]{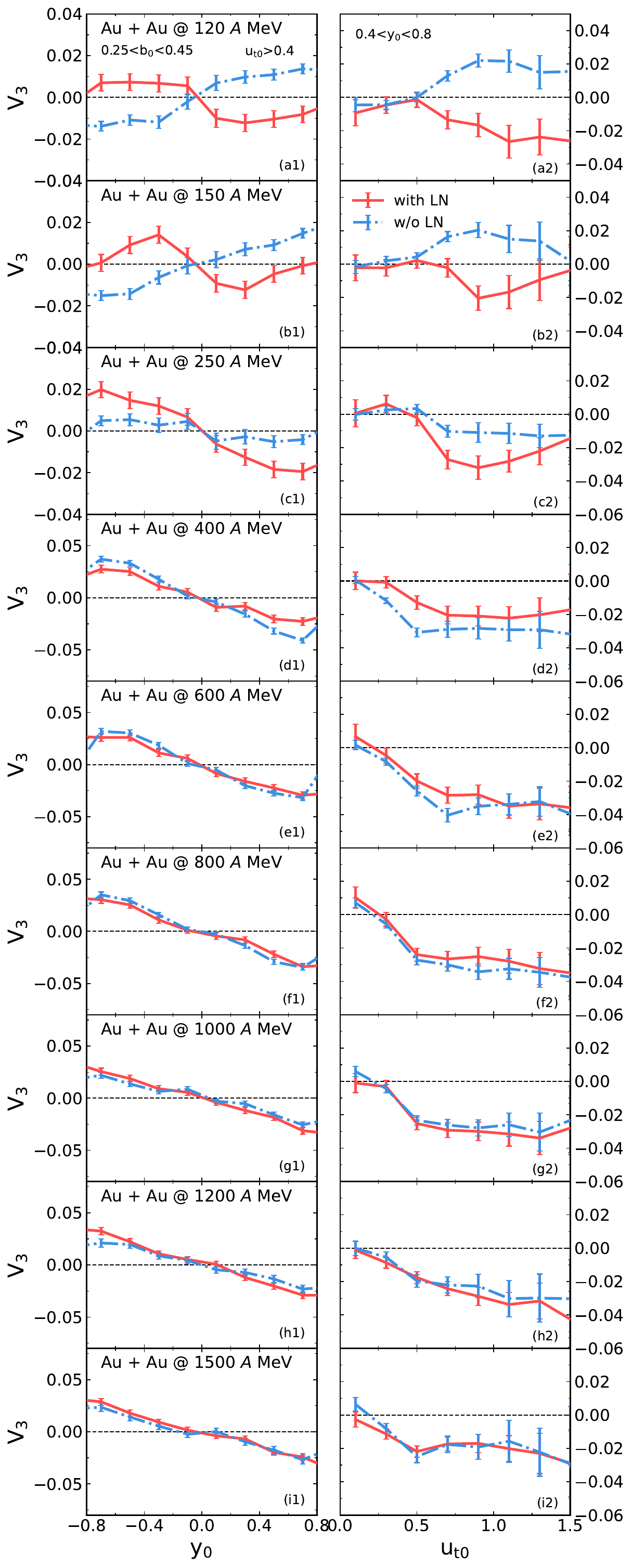}
	\caption{Same as Fig.~\ref{fig:v1energy} but for triangular flow $v_3$.}
	\label{fig:v3energy}
\end{figure}

\begin{figure}[htbp]
	\centering
	\includegraphics[width=0.48\textwidth]{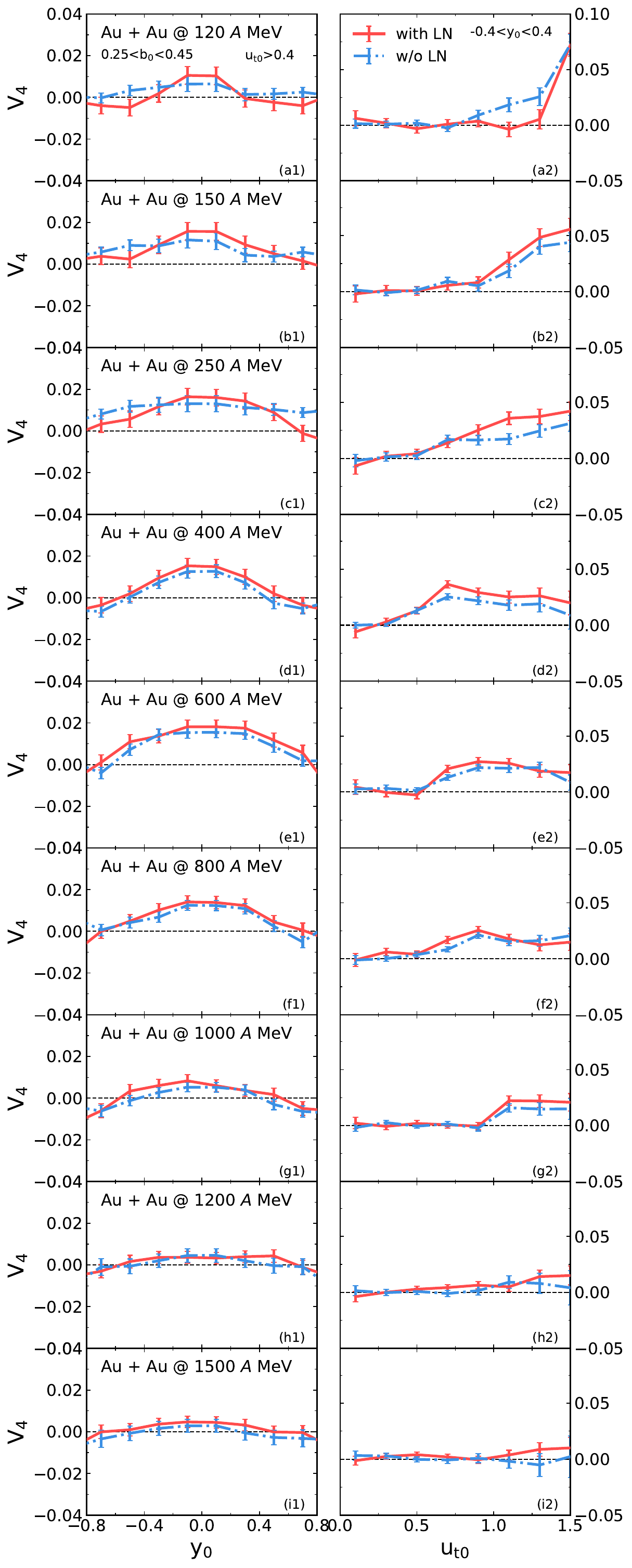}
	\caption{Same as Fig.~\ref{fig:v2energy} but for quadrangular flow $v_4$.}
	\label{fig:v4energy}
\end{figure}

Figures~\ref{fig:v1energy}-\ref{fig:v4energy} show the beam-energy dependence of proton directed ($v_1$), elliptic ($v_2$), triangular ($v_3$), and quadrangular ($v_4$) in Au+Au collisions at $E_{\rm beam}=120$--$1500~A$ MeV with $0.25<b_{0}<0.45$.
In these figures, the cut intervals of the normalized rapidity $y_0$ and the scaled transverse velocity $u_{\rm t0}$ are chosen to be the same as in the FOPI analysis~\cite{FOPI:2011aa}. These quantities are defined as $y_0 = y_z/y_{\rm{pro}}$ with $y_{\rm{pro}}$ being the projectile rapidity in the center-of-mass system, and $u_{\rm{t0}} = u_t/u_{\rm{pro}}$ with $u_t = \beta_t \gamma$ being the transverse component of the four-velocity and $u_{\rm{pro}}$ being the velocity of the incident projectile in the center-of-mass system~\cite{FOPI:2011aa}.
Calculations including explicit light-nucleus degrees of freedom are compared with the corresponding calculations without light nuclei; for brevity, these two cases are labeled as ``with LN'' and ``w/o LN'' in the figures, respectively. In Figs.~\ref{fig:v1energy} and~\ref{fig:v3energy}, the left panels show the normalized rapidity dependence of $v_1$ and $v_3$, respectively, for scaled transverse-velocity $u_{\rm t0}>0.4$, while the right panels show the scaled transverse-velocity dependence in the forward-rapidity region $0.4<y_0<0.8$ for $v_1$ and $v_3$, respectively.
In Figs.~\ref{fig:v2energy} and~\ref{fig:v4energy}, the left panels show the normalized rapidity dependence of $v_2$ and $v_4$, respectively, for $u_{\rm t0}>0.4$, while the right panels show the scaled transverse-velocity dependence in the midrapidity region $-0.4<y_0<0.4$.

One sees from Fig.~\ref{fig:v1energy} that the proton directed flow $v_1$ exhibits a clear energy dependence of the cluster-induced modification. At the two lowest beam energies, $E_{\rm beam}=120$ and $150~A$ MeV, the ``with LN'' calculation shows a significant enhancement in the magnitude of proton $v_1$. In contrast, the ``w/o LN'' calculation gives a weak and rather flat $v_1(y_0)$, especially at $E_{\rm beam}=120~A$ MeV.
The scaled transverse-velocity dependence of proton $v_1$ shows a similar trend: the ``with LN'' calculation gives much larger $v_1$ values at low and intermediate $u_{\rm t0}$, whereas the ``w/o LN'' calculation remains significantly smaller. This indicates that, in the lowest-energy region, the final free-proton flow is strongly affected by the redistribution of baryon number between unbound nucleons and dynamically formed light clusters. At $E_{\rm beam}=250~A$ MeV, the light-cluster effect remains pronounced, although it is weaker than at $E_{\rm beam}=120$ and $150~A$ MeV. The ``with LN'' calculation still gives a noticeably larger $v_1$ magnitude over most of the rapidity range, and the difference between the ``with LN'' and ``w/o LN'' calculations remains visible in the $u_{\rm t0}$ dependence. At $E_{\rm beam}=400~A$ MeV, the two calculations become closer, but the light-cluster effect remains evident, particularly at large $y_0$ and high $u_{\rm t0}$. This suggests that $E_{\rm beam}=250$--$400~A$ MeV represents a transition region in which light-cluster production is still sufficiently abundant to modify the final free-proton phase-space distribution, but the cluster formation effect is already weaker than at $E_{\rm beam}=120$--$150~A$ MeV.
With further increasing beam energy, the difference between the  ``with LN'' and ``w/o LN'' calculations becomes progressively smaller. At $E_{\rm beam}=600~A$ MeV, the modification of $v_1$ is already moderate. From $E_{\rm beam}=800~A$ MeV to $E_{\rm beam}=1500~A$ MeV, the two calculations nearly overlap in both the normalized rapidity and scaled transverse-velocity dependences, with only small residual differences in selected high-$u_{\rm t0}$ or large-$y_{0}$ regions. This behavior is consistent with the decreasing light-cluster abundance and survival probability at higher beam energies, where the final proton collective motion is less affected by cluster formation.

The elliptic flow shown in Fig.~\ref{fig:v2energy} also displays a clear beam-energy dependence of the cluster-induced modification of proton elliptic flow. In the ``w/o LN'' calculation, the magnitude of proton $v_2$ remains rather weak at $E_{\rm beam}=120$ and $150~A$ MeV over a broad normalized rapidity and scaled transverse velocity range. A noteworthy feature appears at $E_{\rm beam}=120$ $A$ MeV, where the proton $v_2$ becomes positive around midrapidity, with positive values also observed in the scaled transverse velocity dependence. This positive elliptic flow indicates that, without light-cluster degrees of freedom, the final free-proton emission at the lowest beam energy of $E_{\rm beam}=120$ $A$ MeV is not dominated by the squeeze-out component, but contains a sizable in-plane emission contribution. In the ``with LN'' calculation, the proton $v_2$ at $E_{\rm beam}=120$ and $150~A$ MeV is driven to more negative values, especially around midrapidity and at larger $u_{\rm t0}$. This change corresponds to an enhanced squeeze-out component in the final free-proton emission. It suggests that light-cluster dynamics does not simply change the final particle composition, but also modifies the phase-space distribution of the remaining free protons. In particular, the redistribution of baryon number between free nucleons and bound light clusters can alter the balance between in-plane and out-of-plane emission—a trend that has been experimentally observed in Refs.~\cite{FOPI:2004bfz,LeFevre:2016vpp}. The change in proton $v_2$ from a weak or even positive value in the ``w/o LN'' calculation to a more negative one in the ``with LN'' calculation demonstrates that proton elliptic flow is especially sensitive to light-cluster dynamics at these lower FOPI energies.
At $E_{\rm beam}=250~A$ MeV, the ``with LN'' calculation still gives a more negative proton $v_2$ than the ``w/o LN'' calculation, especially around midrapidity and at larger $u_{\rm t0}$. At $E_{\rm beam}=400~A$ MeV, the difference becomes smaller but remains visible, indicating that cluster formation still affects the proton elliptic flow in this energy region. At $E_{\rm beam}=600~A$ MeV and above, the results of the two calculations ``with LN'' and ``w/o LN'' are much closer. In particular, from $E_{\rm beam}=800$ to $E_{\rm beam}=1500~A$ MeV, the $v_2(y_0)$ and $v_2(u_{\rm t0})$ results obtained in the ``with LN'' and ``w/o LN'' calculations are very similar, showing that the influence of explicit light-cluster degrees of freedom on proton elliptic flow becomes weak in the higher energy range studied.

The triangular flow $v_3$ shown in Fig.~\ref{fig:v3energy} provides an additional test of the cluster-induced modification of the final proton flows. At $E_{\rm beam}=120$ and $150~A$ MeV, the inclusion of light nuclei changes not only the magnitude but also the sign pattern of $v_3$ in both the normalized rapidity and scaled transverse-velocity dependences. In particular, the ``with LN'' calculation gives a more negative $v_3$ value at forward rapidities and at intermediate-to-large $u_{\rm t0}$, whereas the ``w/o LN'' calculation tends to give positive or much weaker values in the same kinematic regions. This indicates that the inclusion of light-cluster formation also affects higher-order anisotropies. At $E_{\rm beam}=250$ and $400~A$ MeV, the difference between the two calculations remains visible but becomes less pronounced. Above about $600~A$ MeV, the ``with LN'' and ``w/o LN'' results for $v_3$ are much closer over most of the rapidity and transverse-velocity ranges, showing that the cluster effect on proton triangular flow is also reduced at higher beam energies.

The quadrangular flow $v_4$ in Fig.~\ref{fig:v4energy} shows a weaker sensitivity to light-cluster degrees of freedom than $v_1$, $v_2$, and $v_3$. The calculated $v_4$ is generally small, with a positive structure around midrapidity and a mild increase at larger $u_{\rm t0}$ in the low- and intermediate-energy region. The inclusion of light nuclei slightly modifies the magnitude of $v_4$, especially from $E_{\rm beam}=150$ to $600~A$ MeV, but the qualitative rapidity and transverse-velocity dependences remain similar to those obtained without light nuclei. At higher beam energies, the results of the two calculations with and without considering light nuclei are nearly indistinguishable within the displayed uncertainties.

Overall, Figures~\ref{fig:v1energy}-\ref{fig:v4energy} reveal a systematic beam-energy dependence of the light-cluster effects on proton flows. The modification is most pronounced at $E_{\rm beam}=120$--$150~A$ MeV, remains appreciable at $E_{\rm beam}=250~A$ MeV, becomes moderate around $E_{\rm beam}=400~A$ MeV, and is gradually reduced from $E_{\rm beam}=600~A$ to $1500~A$ MeV. This trend can be connected with the beam-energy dependence of the light-nucleus yields as discussed in Sec.~\ref{LNyield}. In the low-energy region, the larger light-cluster abundance implies a stronger redistribution of baryon number from free nucleons into bound clusters, thereby reshaping the final free-proton phase-space distribution and its collective flow.  With increasing beam energy, the decreasing light-cluster abundance and reduced survival probability of clusters weaken their influence on proton flows. It should also be emphasized that the cluster-induced modification is not simply an overall enhancement or suppression of proton flow. Its sign and magnitude depend on the beam energy as well as on the selected rapidity and transverse-velocity region. At the same time, the reduced difference between the ``with LN'' and ``w/o LN'' calculations at higher beam energies further suggests that light-cluster degrees of freedom have only a limited effect on the calculated proton flow in the higher-energy region.

\begin{figure*}[htbp]
	\centering
	\includegraphics[width=0.8\textwidth]{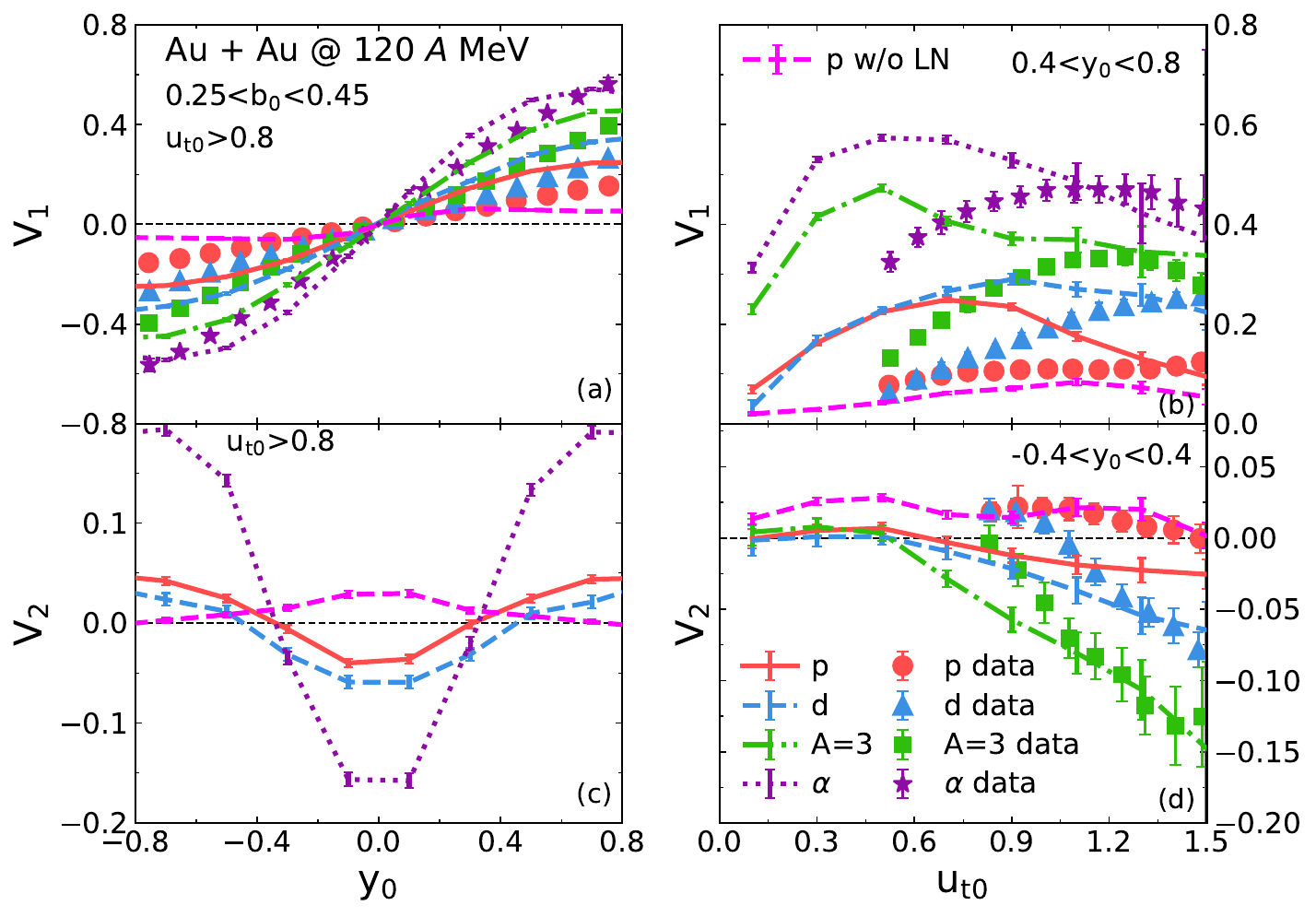}
	\caption{Directed ($v_1$) and elliptic ($v_2$) flows as functions of normalized rapidity $y_{0}$ and scaled transverse velocity $u_{\rm{t0}}$ for protons and light nuclei in Au+Au collisions at $E_{\rm{beam}}=120$ $A$ MeV. The lines are predictions from the lattice BUU transport model. The corresponding FOPI data~\cite{FOPI:2011aa} are also included for comparison.}
	\label{fig:120rap}
\end{figure*}

\begin{figure*}[htbp]
	\centering
	\includegraphics[width=0.8\textwidth]{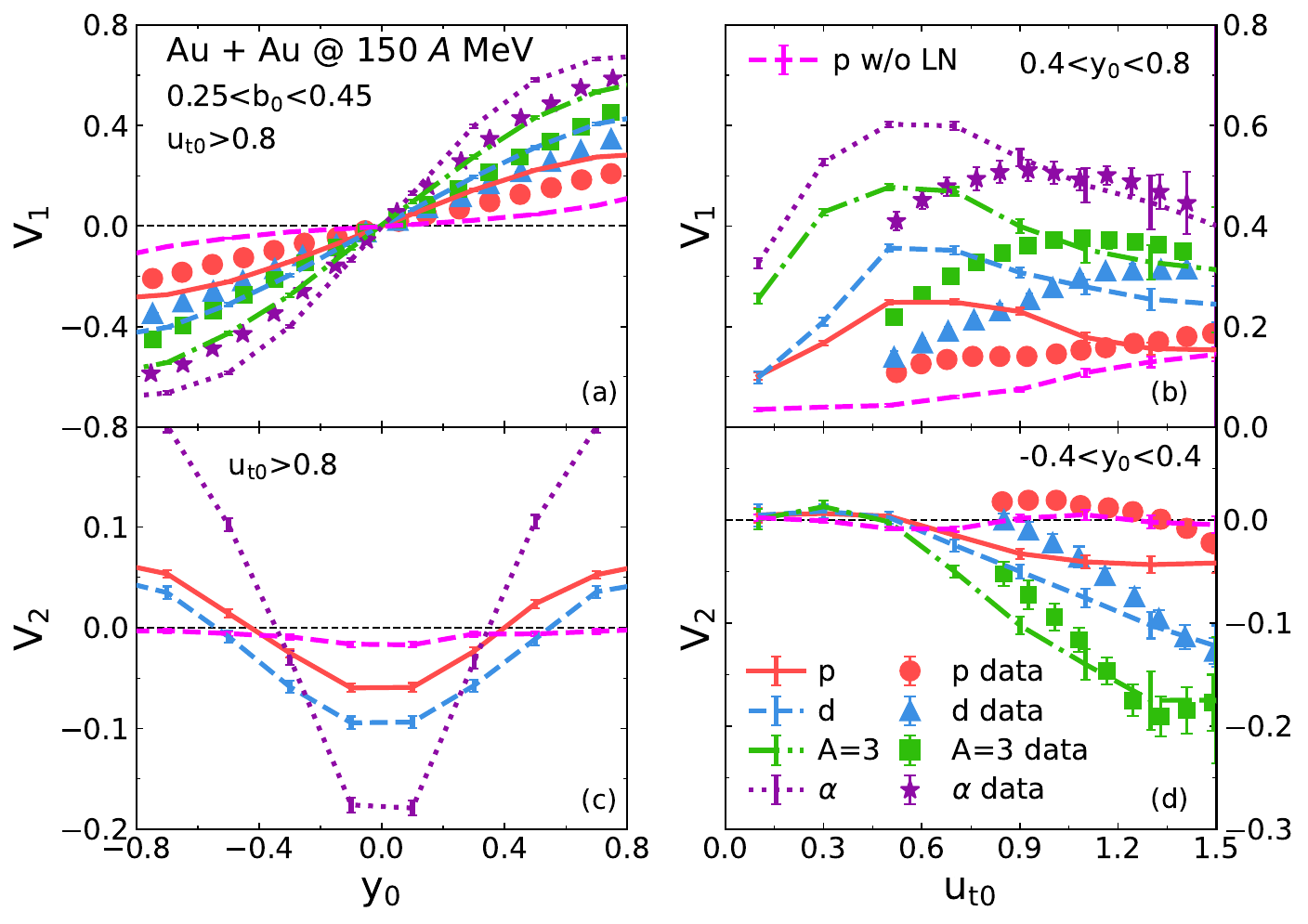}
	\caption{Same as Fig.~\ref{fig:120rap} but for Au+Au collisions at $E_{\rm{beam}}=$150$$ $A$ MeV.}
	\label{fig:150rap}
\end{figure*}

\begin{figure*}[htbp]
	\centering
	\includegraphics[width=0.8\textwidth]{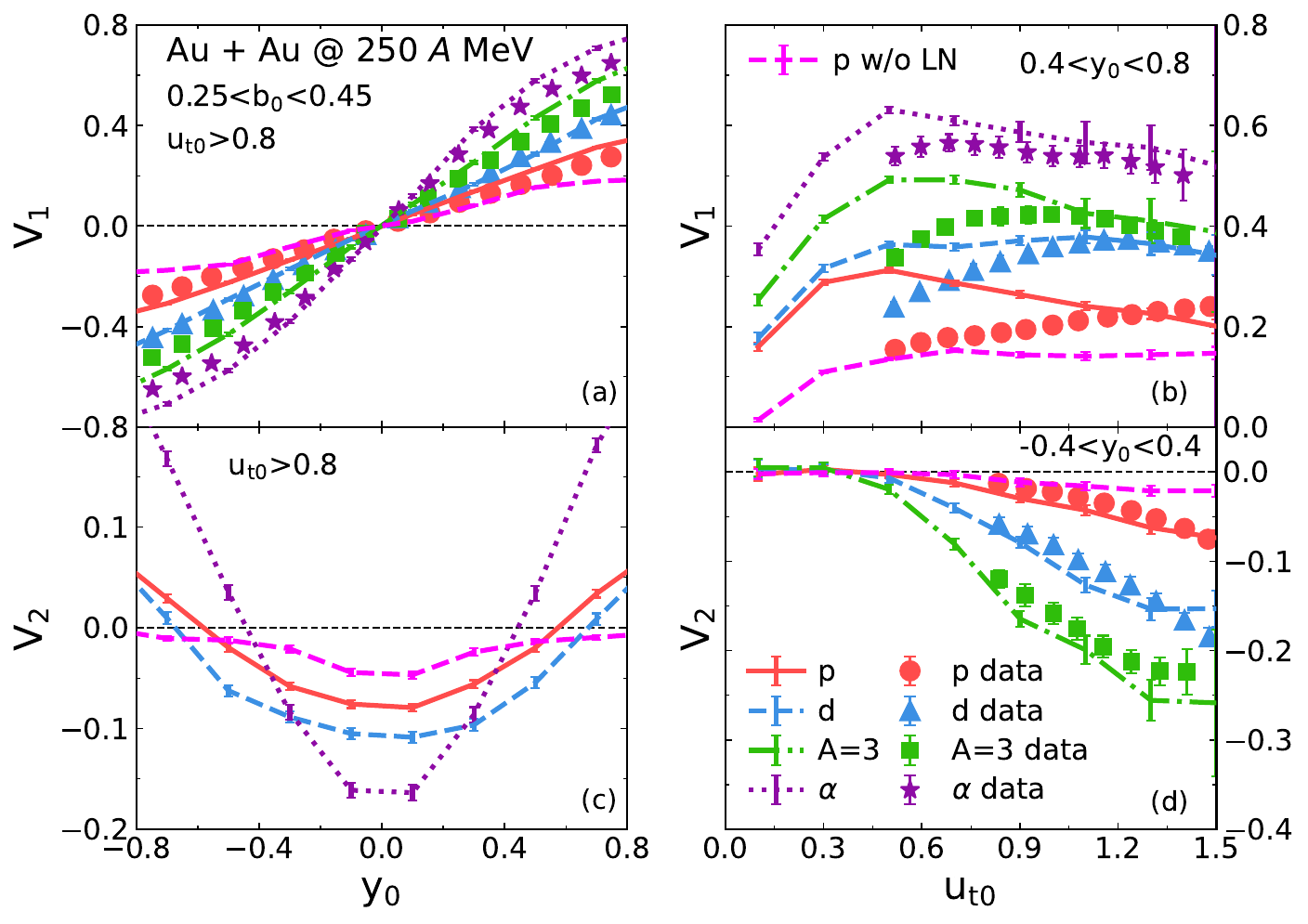}
	\caption{Same as Fig.~\ref{fig:120rap} but for Au+Au collisions at $E_{\rm{beam}}=$250$$ $A$ MeV.}
	\label{fig:250rap}
\end{figure*}

\begin{figure*}[htbp]
	\centering
	\includegraphics[width=0.8\textwidth]{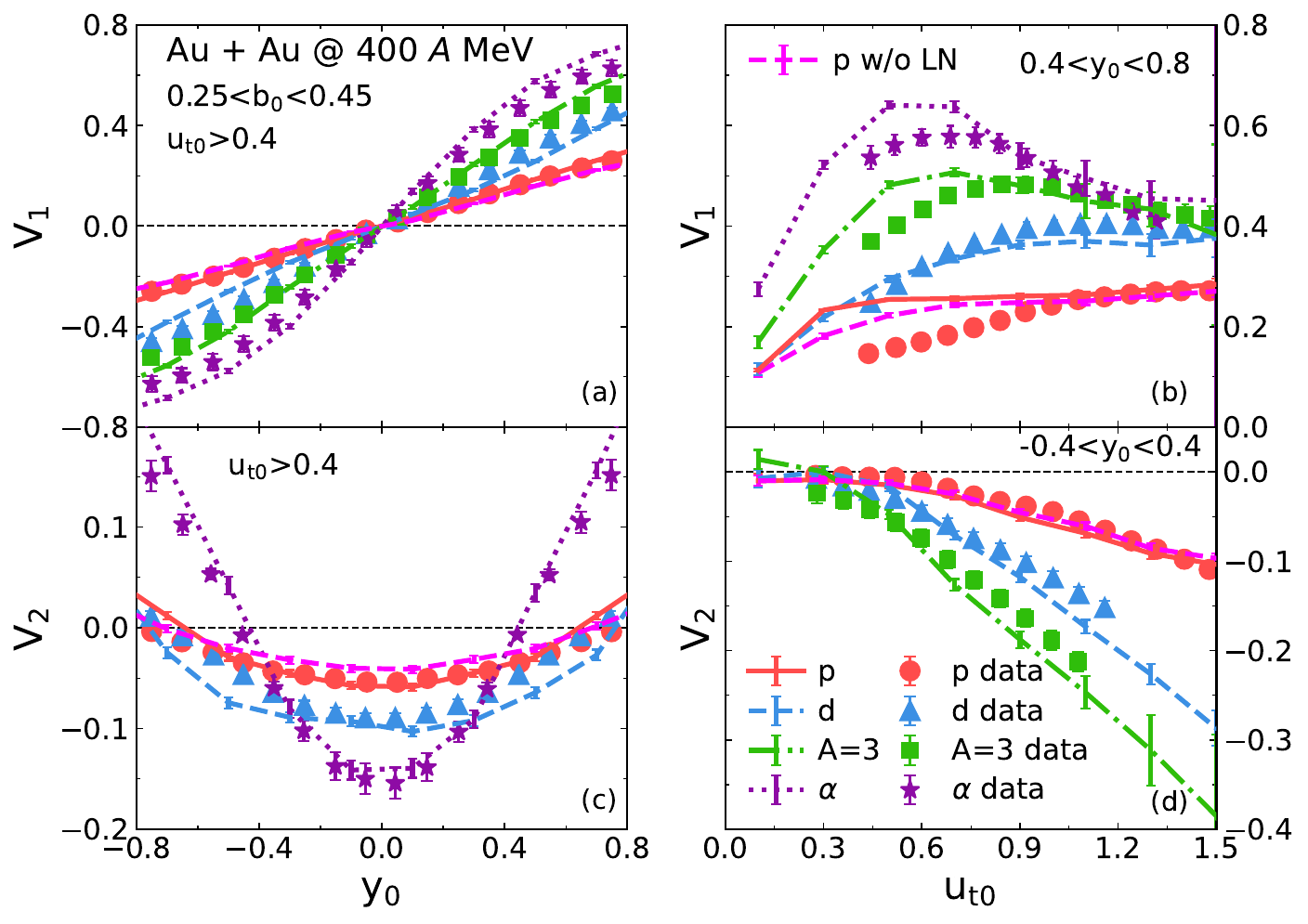}
	\caption{Same as Fig.~\ref{fig:120rap} but for Au+Au collisions at $E_{\rm{beam}}=$400$$ $A$ MeV.}
	\label{fig:400rap}
\end{figure*}

\begin{figure*}[htbp]
	\centering
	\includegraphics[width=0.8\textwidth]{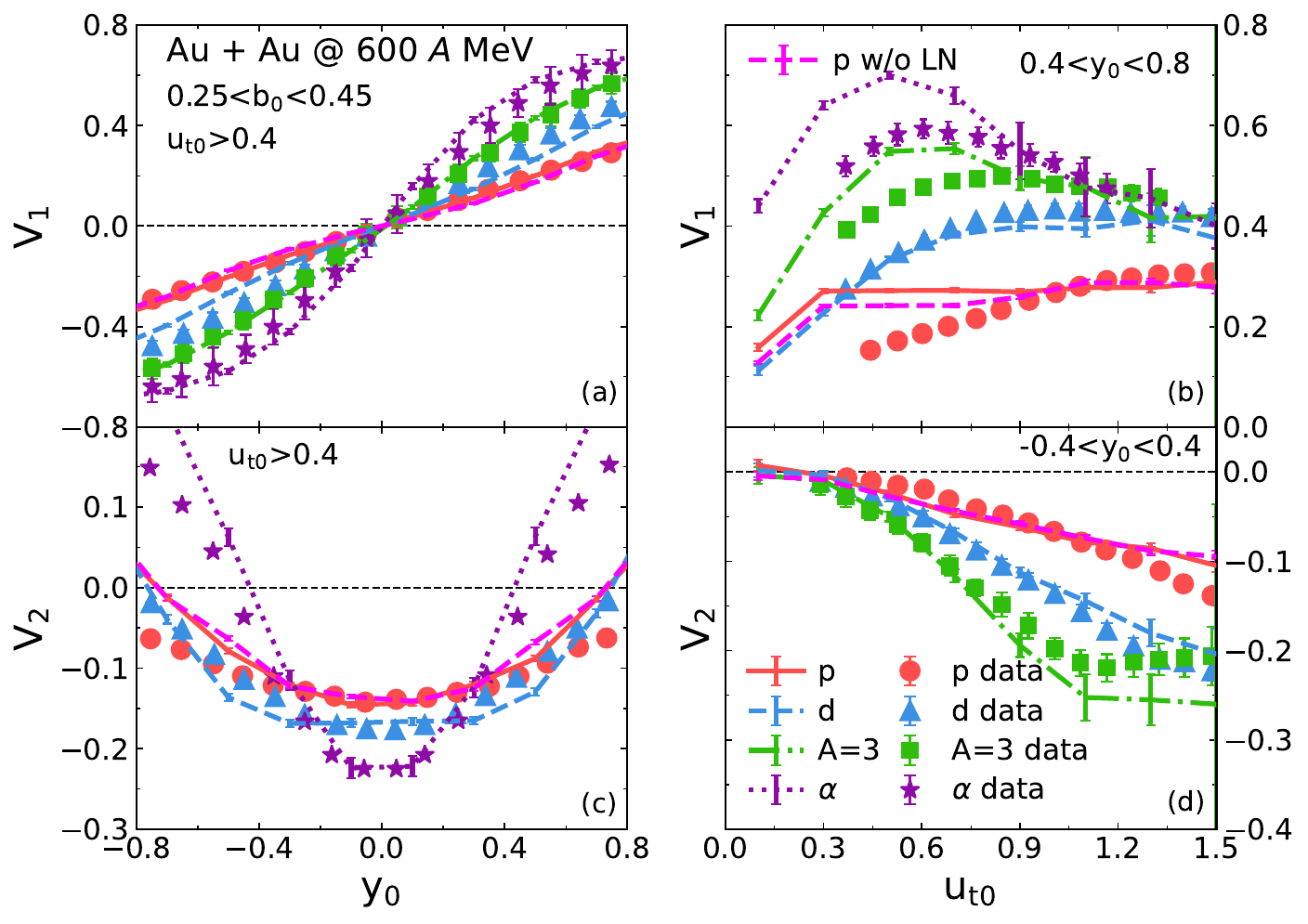}
	\caption{Same as Fig.~\ref{fig:120rap} but for Au+Au collisions at $E_{\rm{beam}}=$600$$ $A$ MeV.}
	\label{fig:600rap}
\end{figure*}

\begin{figure*}[htbp]
	\centering
	\includegraphics[width=0.8\textwidth]{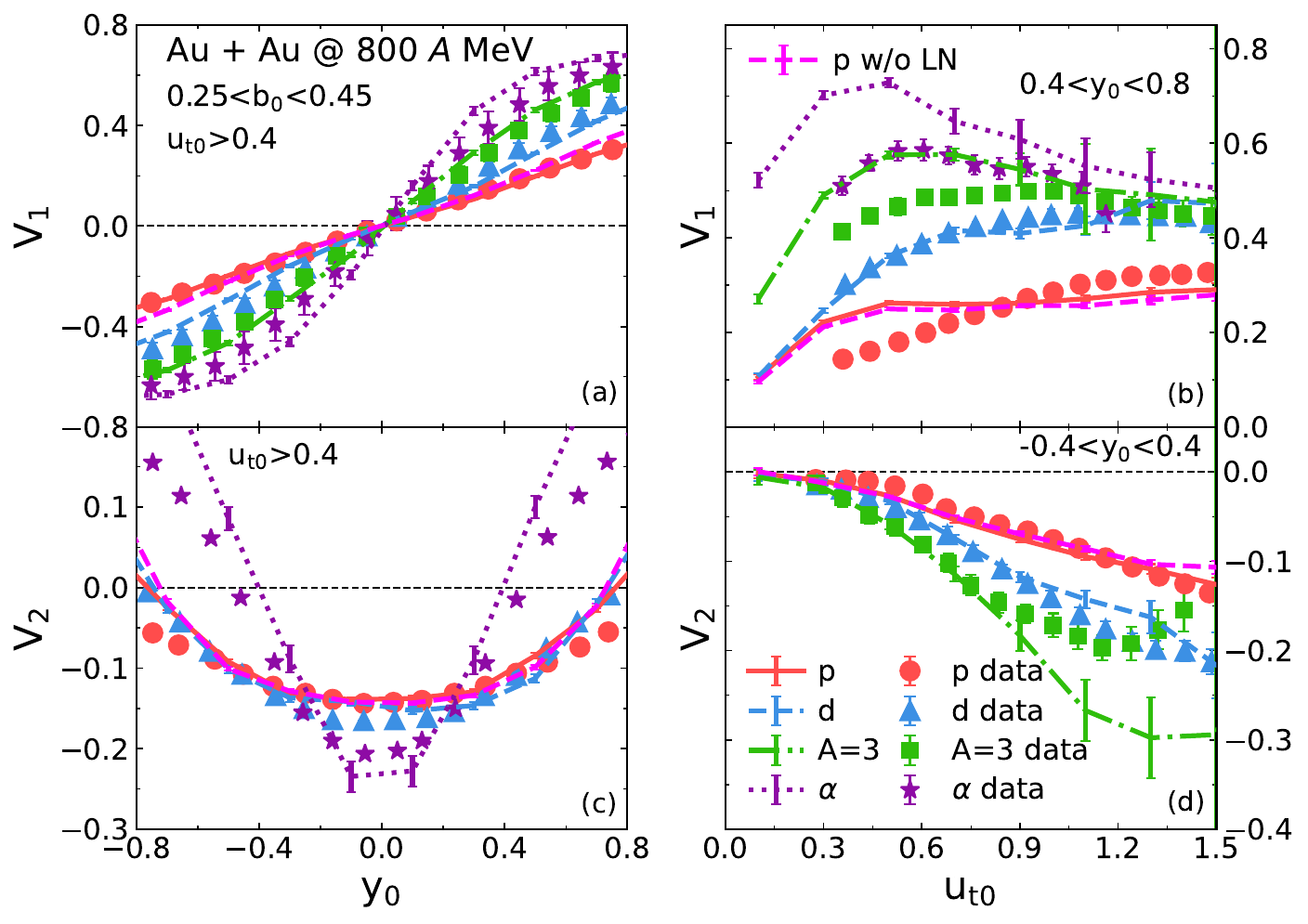}
	\caption{Same as Fig.~\ref{fig:120rap} but for Au+Au collisions at $E_{\rm{beam}}=$800$$ $A$ MeV.}
	\label{fig:800rap}
\end{figure*}

\begin{figure*}[htbp]
	\centering
	\includegraphics[width=0.8\textwidth]{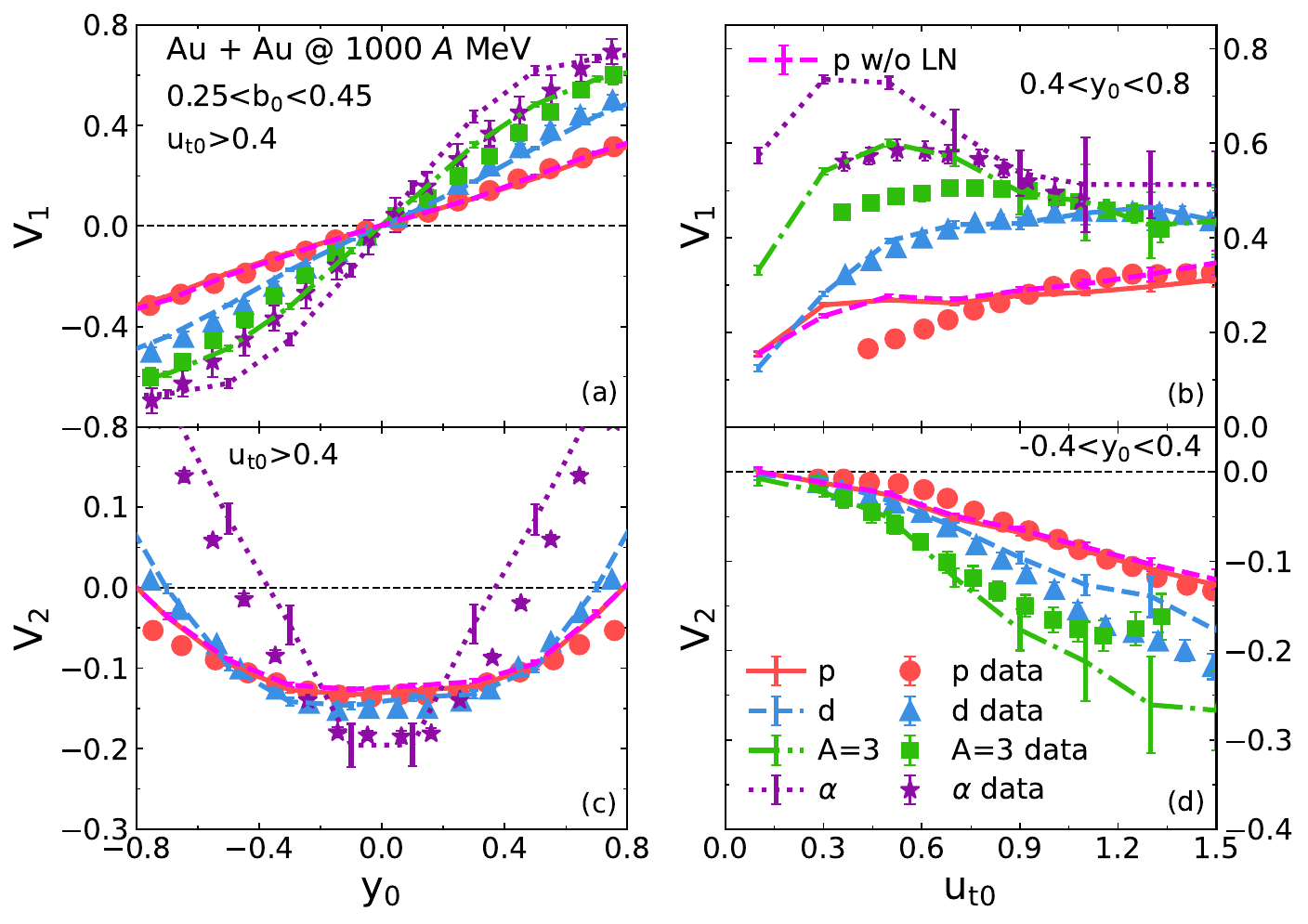}
	\caption{Same as Fig.~\ref{fig:120rap} but for Au+Au collisions at $E_{\rm{beam}}=$1000$$ $A$ MeV.}
	\label{fig:1000rap}
\end{figure*}

\begin{figure*}[htbp]
	\centering
	\includegraphics[width=0.8\textwidth]{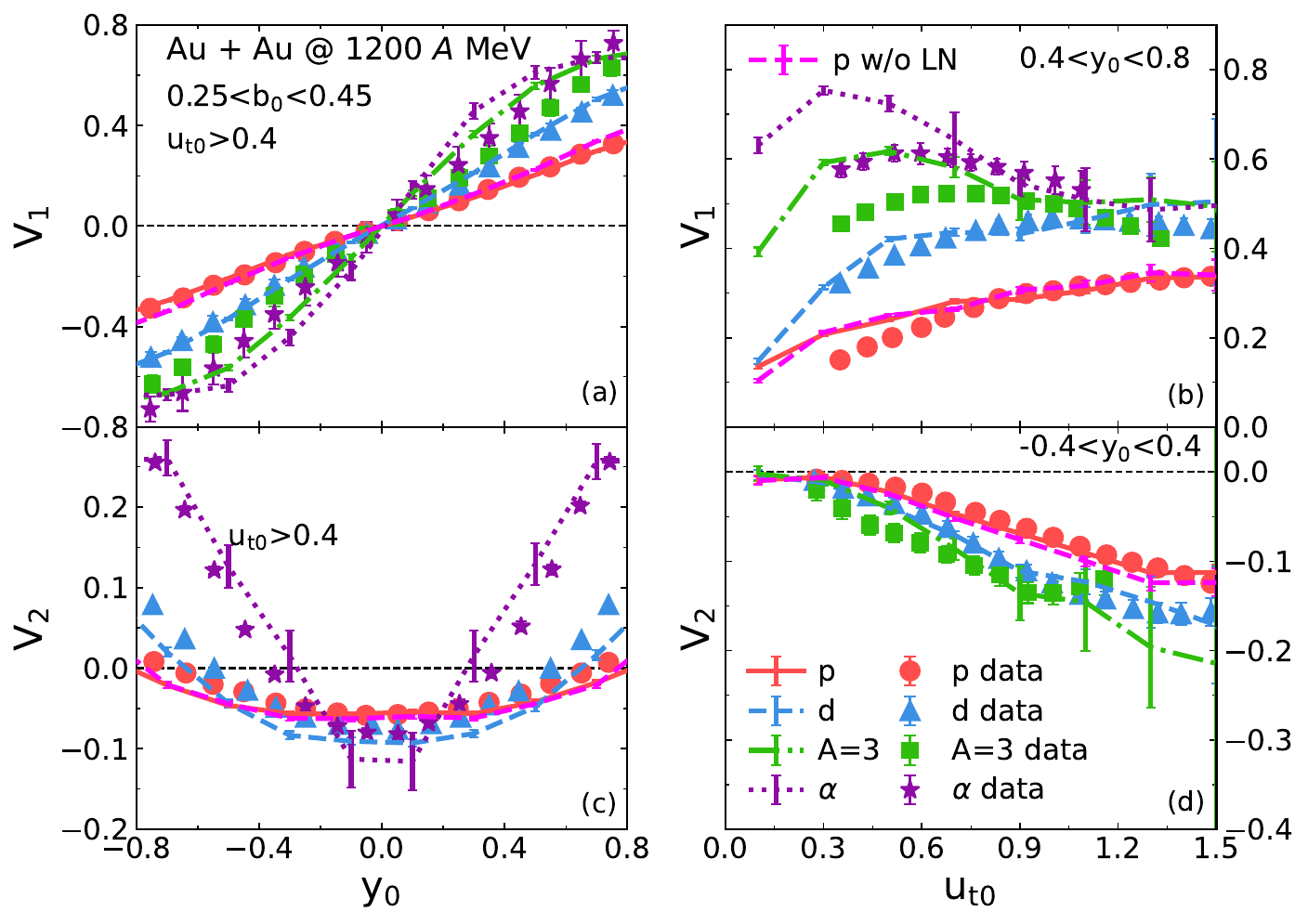}
	\caption{Same as Fig.~\ref{fig:120rap} but for Au+Au collisions at $E_{\rm{beam}}=$1200$$ $A$ MeV.}
	\label{fig:1200rap}
\end{figure*}

\begin{figure*}[htbp]
	\centering
	\includegraphics[width=0.9\textwidth]{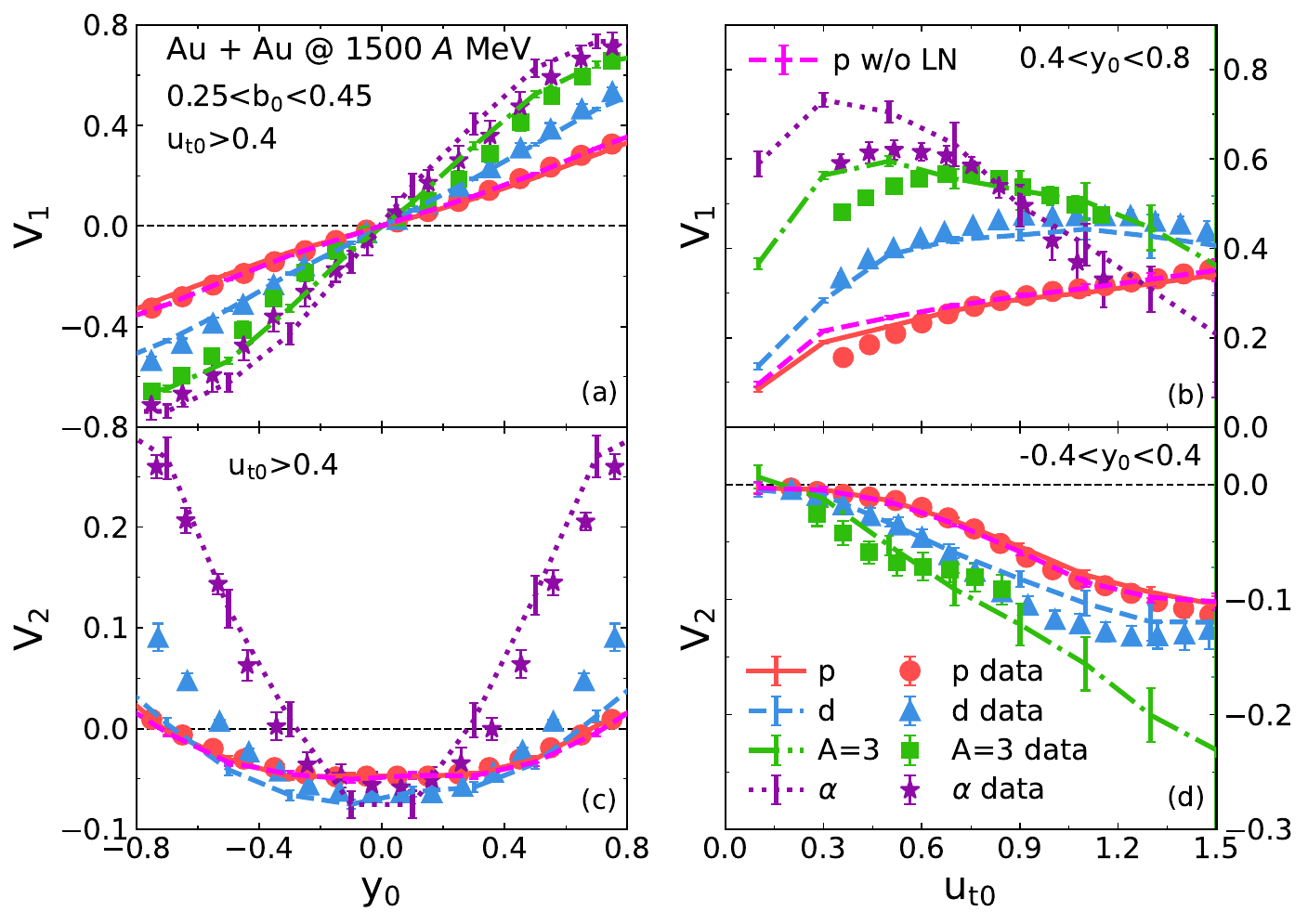}
	\caption{Same as Fig.~\ref{fig:120rap} but for Au+Au collisions at $E_{\rm{beam}}=$1500$$ $A$ MeV.}
	\label{fig:1500rap}
\end{figure*}

\subsection{Collective flows of protons and light nuclei from  $E_{\rm{beam}}$= $120~A$ to $1500~A$ MeV}

We next examine whether the kinetic approach for light-nucleus formation can describe the collective flows of protons and light nuclei over the studied beam-energy range. Figures~\ref{fig:120rap}--\ref{fig:1500rap} compare the calculated directed flow $v_1$ and elliptic flow $v_2$ with the FOPI data~\cite{FOPI:2011aa} for Au+Au collisions at $E_{\rm beam}=120$--$1500~A$ MeV. For the normalized rapidity dependences, a cut of $u_{\rm t0}>0.8$ is applied at $E_{\rm beam}=120$--$250~A$ MeV, while $u_{\rm t0}>0.4$ is used at $E_{\rm beam}=400$--$1500~A$ MeV. The scaled transverse-velocity dependences are shown in the forward-rapidity region $0.4<y_0<0.8$ for $v_1$ and at midrapidity, $-0.4<y_0<0.4$, for $v_2$. The ``w/o LN'' proton calculation is also shown as a reference to further illustrate how dynamical light-cluster formation modifies the final free-proton flow.

At the lower beam energies, $E_{\rm beam}=120$, $150$, and $250$ $A$ MeV, the present kinetic approach for light-nucleus formation provides only a partial description of the experimental flow data. For protons, the comparison between the ``with LN'' and ``w/o LN'' calculations shows that dynamical light-cluster formation has a strong impact on both $v_1$ and $v_2$, as discussed in Sec.~\ref{energyflow}. For light nuclei, however, the calculated flow magnitudes tend to be larger than the measured values, especially in selected large-rapidity regions, and this overestimation occurs at low $u_{\rm t0}$ for $v_1$ and at high $u_{\rm t0}$ for $v_2$. This indicates that the present calculation overestimates the collectivity carried by dynamically produced light clusters in the low-energy region.
The possible origin of the low-energy discrepancy may be related to the absence of explicit treatment of heavier fragments,residual nuclei, and inter-cluster correlations, as well as uncertainties in the in-medium suppression of light clusters as discussed in Secs.~\ref{LNyield} and~\ref{energyflow}.
The large-rapidity region in semi-central collisions may also be affected by spectator fragmentation or by the formation of heavier fragments associated with projectile- and target-like remnants. Such effects could also modify the forward-rapidity flow, including the scaled transverse velocity dependence of $v_1$, but they are not explicitly included in the present dynamical calculation. Therefore, the low-energy comparison indicates that the kinetic approach for light-nucleus formation captures an important part of the flow systematics, but is not sufficient by itself for a quantitative description of light-nucleus flow observables at $E_{\rm beam}=120$--$250$ $A$ MeV.

For $E_{\rm beam}\geq 400~A$ MeV, the overall description of the experimental flow data becomes more satisfactory. The kinetic approach for light-nucleus formation provides an overall reasonable description of the rapidity and scaled transverse-velocity dependences of proton and light-nucleus $v_1$ and $v_2$ over the beam-energy range from $400$ to $1500~A$ MeV.
Nevertheless, some differential discrepancies remain. For $v_1$, a visible discrepancy with the data remains at the low-$u_{\rm t0}$ region, especially at $E_{\rm beam}=400$--$800~A$ MeV. These deviations may be related to spectator-fragmentation effects at forward rapidities, and could also be related to the treatment of in-medium elastic scattering, nucleon--cluster and cluster--cluster interactions, possible medium modifications of light clusters, and the propagation and breakup of composite clusters in the evolving medium. For $v_2$, discrepancies remain mainly at large $y_{0}$ and high $u_{\rm t0}$, where the calculated light-nucleus elliptic flows do not fully reproduce the measured values. This might be attributed to the interplay of the correlations between light clusters and surrounding nucleons, as well as to the mean-field potentials and in-medium propagation of light clusters.

Overall, Figs.~\ref{fig:120rap}--\ref{fig:1500rap} demonstrate that the kinetic approach for light-nucleus formation provides an overall reasonable description of proton and light-nucleus collective flows over the studied beam-energy range, thereby highlighting the importance of treating light nuclei as dynamical degrees of freedom in flow analyses. At the same time, the limitations are most evident at $E_{\rm beam}=120$--$250~A$ MeV, where the light-nucleus flow magnitudes are generally overpredicted. This discrepancy indicates the possible importance of heavier fragments, spectator remnants, and refined in-medium cluster properties in the low-energy region. At $E_{\rm beam}\geq 400~A$ MeV, the description becomes more satisfactory, although remaining deviations for light nuclei, e.g., at low $u_{\rm t0}$ in $v_1$ and at large $y_{0}$ and high $u_{\rm t0}$ in $v_2$, suggest that cluster interactions, correlations, and in-medium propagation still require further constraints. Importantly, we would like to point out that despite these residual discrepancies in the light-nucleus flows, the model provides an overall satisfactory description of the FOPI proton $v_1$ and $v_2$ data across the full beam-energy range considered.

\begin{figure}[htbp]
	\centering
	\includegraphics[width=0.37\textwidth]{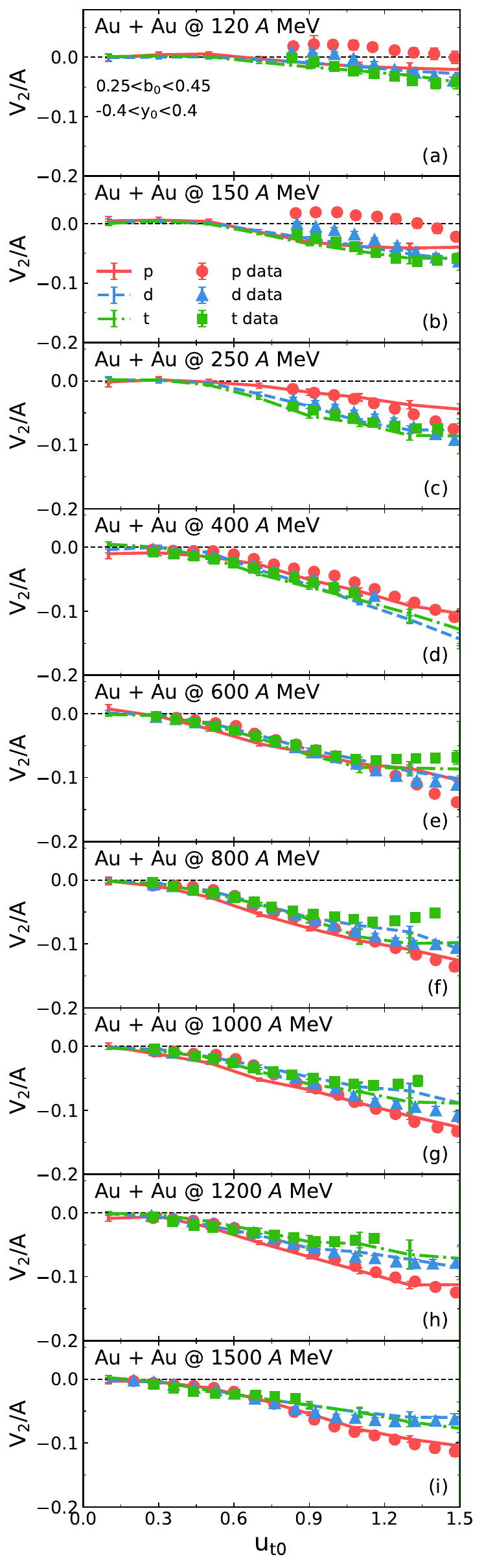}
	\caption{Nucleon-number scaled elliptic flow ($v_2$/A) as a function of scald transverse velocity $u_{\rm{t0}}$ for protons ($p$), deuterons ($d$), and tritons ($t$) in Au+Au collisions at beam energies ranging from $120$--$1500$ $A$ MeV.}
	\label{fig:1500400scaling}
\end{figure}

\subsection{Nucleon-number scaling of light-nucleus elliptic flow}
Having discussed the collective flows of protons and light nuclei, we further examine the nucleon-number scaling of the elliptic flow. This scaling provides a useful test of how the collective motion of a composite light nucleus is related to that of its constituent nucleons. In a simple late-stage coalescence picture~\cite{Kolb:2004gi}, if a light nucleus is formed from nucleons that are close in phase space and if the nucleon flow varies smoothly over the coalescence region, one may expect an approximate relation $v_2^A \simeq A v_2^N$~\cite{Yan:2006bx,Wang:2019eec}. In such a case, the scaled elliptic flow $v_2/A$ of different particle species would approximately follow a common trend when plotted as a function of the corresponding scaled transverse velocity. Since light nuclei are treated as dynamical degrees of freedom in the present kinetic approach, it is important to test whether this simple scaling expectation is preserved.

Figure~\ref{fig:1500400scaling} shows the nucleon-number-scaled elliptic flow $v_2/A$ as a function of $u_{\rm t0}$ for protons, deuterons and tritons from $E_{\rm beam}=120$ to $1500~A$ MeV, comparing the model calculations with the FOPI data. Since there are no data for $^{3}\mathrm{He}$ and $^{4}\mathrm{He}$ particles, we do not show their results in Fig.~\ref{fig:1500400scaling}. At the lower beam energies, especially for $E_{\rm beam}= 120-400$ $A$ MeV, the $v_2/A$ values of deuterons and tritons does not collapse onto a common curve with that of protons over the full $u_{\rm t0}$ range. The deviation becomes more visible at larger $u_{\rm t0}$, where the light-nucleus $v_2/A$ separates from the proton trend. This behavior indicates that the elliptic flow of deuterons and tritons in this energy region cannot be described as a simple sum of independent nucleons at the final stage.
The deviation from nucleon-number scaling in the low-energy region may reflect not only the dynamical effects in light-cluster formation, but also missing ingredients such as contributions from heavier fragments, spectator fragmentation in semi-central collisions, and uncertainties in the in-medium suppression and propagation of composite clusters.

With increasing beam energy, the scaling behavior becomes closer to the nucleon-number expectation in limited kinematic regions. From $E_{\rm beam}=600$ to $1500~A$ MeV, the $v_2/A$ values of protons, deuterons, and tritons are relatively close at low and intermediate $u_{\rm t0}$. This suggests that, at higher beam energies, the elliptic flow of light nuclei becomes more consistent with a nucleon-coalescence-like behavior, in which the collective motion of a cluster is more closely connected to that of its constituent nucleons. Nevertheless, noticeable deviations remain at larger $u_{\rm t0}$, implying that high $u_{\rm t0}$  clusters still probe selected regions of phase space and retain information on the microscopic formation and propagation dynamics. The residual scaling violation at large $u_{\rm t0}$ may have a dynamical origin related to the formation time and in-medium survival of fast-moving light nuclei. High-$u_{\rm t0}$ clusters may be preferentially produced at earlier stages of the reaction and subsequently escape more rapidly from the dense interaction region. Although the medium is denser at early times, the Mott suppression is momentum dependent. For a cluster with a large center-of-mass momentum, the momenta of its constituent nucleons may have a smaller overlap with the highly occupied low-momentum states of the surrounding medium, thereby reducing the averaged phase-space occupation entering Eq.~(\ref{E:fcut})~\cite{Ropke:1982ino,Ropke:1983lbc,Wang:2025lsm}. Such clusters may therefore experience a weaker effective Mott suppression, while their rapid escape reduces the probability of subsequent breakup and rescattering. As a result, high-$u_{\rm t0}$ light nuclei may have a larger survival probability and retain flow information from an earlier and more restricted stage of the collision. Their elliptic flow would then not necessarily follow the late-stage coalescence-like scaling observed at low and intermediate $u_{\rm t0}$. This mechanism provides a possible explanation for the persistent scaling deviations at large $u_{\rm t0}$ even at higher beam energies.

The nucleon-number scaling result should therefore not be interpreted as uniquely identifying light-cluster formation mechanism. Rather, it demonstrates that a purely kinematic coalescence scaling is insufficient to describe the full differential behavior of the calculated collective flows.
Combined with the reasonable description of the species dependence in Fig.~\ref{fig:1500400scaling}, the scaling analysis supports the need for a dynamical treatment in which light nuclei are coupled to the transport evolution.
At the same time, we want to point out that different microscopic light-cluster formation mechanisms, including final-state coalescence, kinetic production through reaction dynamics, and cluster formation from many-body correlations, may lead to different behaviors of scaling while still giving comparable descriptions of more inclusive light-cluster observables. More differential observables, such as the correlation functions involving nucleon--cluster and cluster--cluster pairs, are therefore essential for distinguishing among these scenarios.

\section{Summary and Outlook}
\label{summary}

In this work, we have investigated the role of explicit light-cluster degrees of freedom in collective-flow observables in intermediate-energy Au+Au collisions at FOPI energies. The calculations are performed within the LBUU transport model supplemented by a kinetic light-nucleus formation mechanism, in which deuterons, tritons, $^{3}\mathrm{He}$, and $^{4}\mathrm{He}$ are formed, dissociated, scattered, and propagated dynamically together with nucleons and other hadronic degrees of freedom.
We first benchmark the kinetic approach with FOPI light-nucleus yields in central ($b_{0}<0.15$) collisions over $E_{\rm beam}=120$--$1500~A$ MeV, then examined the beam-energy dependence of proton directed, elliptic, triangular, and quadrangular flows by comparing calculations with and without light clusters in mid-central ($0.25<b_{0}<0.45$) collisions over the same beam-energy range. We further compare the calculated flows of protons and light nuclei with the available FOPI flow data, and finally analyze the nucleon-number scaling behavior of light-nucleus elliptic flow.

The light-nucleus yield comparison shows that the kinetic approach gives a reasonable description of the overall beam-energy dependence of deuteron, triton, $^{3}\mathrm{He}$, and $^{4}\mathrm{He}$ production. However, the light-cluster yields are overestimated at the lowest beam energies, indicating that the low-energy region may require a more complete treatment of heavier fragments, residual nuclei, inter-cluster correlations, and in-medium cluster suppression.
For proton collective flow, the inclusion of light clusters leads to a clear beam-energy-dependent modification. The effect is strongest at $E_{\rm beam}=120$--$150~A$ MeV, remains visible at $250$--$400~A$ MeV, and becomes weak above about $600$--$1500~A$ MeV. This trend is consistent with the decreasing abundance and survival probability of light clusters at higher beam energies.

For the collective flows of protons and light nuclei, the present calculation gives a comprehensive description of the normalized rapidity and scaled transverse velocity dependences of $v_1$ and $v_2$ over the FOPI energy range. The agreement is less satisfactory at $E_{\rm beam}=120$--$250~A$ MeV, where light-nucleus flow magnitudes are generally overpredicted. At $E_{\rm beam}\geq400~A$ MeV, the description becomes more satisfactory, although residual deviations remain, especially at low $u_{\rm t0}$ for $v_1$ and at large $y_{0}$ and high $u_{\rm t0}$ for $v_2$. These results suggest that differential flow observables provide important constraints on the formation, propagation, and interaction of light clusters in the medium.

The analysis of nucleon-number scaling further shows that the elliptic flow of light nuclei cannot be fully described by a simple coalescence-like scaling picture. Clear scaling violations appear at lower beam energies, especially around $120$--$400~A$ MeV, while the scaling behavior of protons, deuterons, and tritons becomes closer to the nucleon-number expectation at higher beam energies, although visible deviations remain at larger scaled transverse velocity.

In summary, these results demonstrate that explicit light-cluster degrees of freedom are important for a consistent interpretation of collective-flow observables in intermediate-energy heavy-ion collisions, especially in the lower-energy region where cluster production is abundant. However, at higher beam energies ($E_{\rm beam}\geq 600~A$), the proton flows display minor sensitivity to the light-cluster degrees of freedom in heavy-ion collisions. Taken together, the beam-energy-dependent sensitivity of proton flow to explicit light-cluster degrees of freedom, the light-nucleus flow systematics, and the deviations from nucleon-number scaling provide constraints on light-cluster formation and in-medium dynamics that complement those obtained from integrated yields. Future work should extend the present analysis to additional observables, collision systems, and beam energies, and should quantify the uncertainties associated with the nuclear matter equation of state, in-medium modification, Mott-like suppression, and the dynamical treatment of both light clusters and heavier fragments.

\begin{acknowledgments}
This work was supported in part by the National Natural Science Foundation of China under Grant Nos. 12235010, 12575139 and 12147101, the National SKA Program of China (Grant No. 2020SKA0120300), the Science and Technology Commission of Shanghai Municipality (Grant No. 23JC1402700), and the Natural Science Foundation of Henan Province (Grant No. 242300421048).
\end{acknowledgments}

\bibliography{FOPILNflow.bib}

\end{document}